\documentclass{aa}
\input epsf
\begin{document}
\thesaurus{3  
	   (11.09.1  
	    11.16.1  
	    11.19.2  
	    11.19.4  
	   )}
\title{Young massive star clusters in nearby galaxies
  \thanks{Based on observations made with the Nordic Optical Telescope,
          operated on the island of La Palma jointly by Denmark, Finland,
          Iceland, Norway, and Sweden, in the Spanish Observatorio del 
          Roque de los Muchachos of the Instituto de Astrofisica de Canarias,
          and with the Danish 1.5-m telescope at ESO, La Silla, Chile.}
}
\subtitle{I. Identification and general properties of the cluster systems}
\author{S.S. Larsen  \inst{1}
        \and T. Richtler \inst{2}}

\offprints{S.S. Larsen}

\institute{Copenhagen University Astronomical Observatory, 
	   Juliane Maries Vej 32, 2100 Copenhagen {\O}, Denmark \\
	   email: soeren@astro.ku.dk
        \and Sternwarte der Universit{\"a}t Bonn,  
	     Auf dem H{\"u}gel 71, D-53121 Bonn, Germany \\
	   email: richtler@astro.uni-bonn.de}

\date{Received ...; accepted ...}

\maketitle
\markboth{Young massive star clusters..}{}

\begin{abstract}
  Using ground-based UBVRI$H\alpha$ CCD photometry we have been carrying out 
a search for young massive star clusters (YMCs) in a sample consisting of 
21 nearby spiral galaxies. We find a large variety concerning the richness of 
the cluster systems, with some galaxies containing no YMCs at all and others 
hosting very large numbers of YMCs. Examples of galaxies with poor cluster
systems are NGC~300 and NGC~4395, while the richest cluster systems are found 
in the galaxies NGC~5236 (M83), NGC~2997 and NGC~1313. The age distributions of
clusters in these galaxies show no obvious peaks, indicating that massive 
clusters are formed as an ongoing process rather than in bursts. This is in 
contrast to what is observed in starbursts and merger galaxies.  The radial 
distributions of clusters follow the H$\alpha$ surface brightnesses. For the 
galaxies in our sample there is no correlation between the morphological type 
and the presence of YMCs.
\end{abstract}

\section{Introduction}
  
  During the last decade many investigations have revealed the presence
of ``young massive star clusters'' (YMCs) or ``super star clusters'' in 
mergers and starburst galaxies, 
and it has been speculated that these objects could be young analogues of 
the globular clusters seen today in the Milky Way. It is an intriguing
idea that globular clusters could still be formed today in some
environments, because the study of such objects would be expected to provide 
a direct insight into the conditions that were present in the early days of 
our own and other galaxies when the globular clusters we see today in the
halos were formed.

  Probably the most famous example of a merger galaxy hosting ``young massive
clusters'' is the ``Antennae'', NGC~4038/39 where Whitmore \& Schweizer 
(1995) discovered more than 700 blue point-like sources with absolute visual 
magnitudes up to $M_V = -15$. Other well-known examples are NGC~7252 
(Whitmore et. al. 1993), NGC~3921 (Schweizer et. al. 1996) and NGC~1275 
(Holtzman et. al. 1992). All of these galaxies are peculiar systems and 
obvious mergers. In fact, in {\it all} cases investigated so far where star 
formation is associated with a merger, YMCs have been identified 
(Ashman \& Zepf 1998). 

  But YMCs exist not only in mergers. They have been located also in
starburst galaxies such as NGC~1569 and NGC~1705 (O'Connell et. al. 1994),
NGC~253 (Watson et. al. 1996) and M82 (O'Connell et. al. 1995),
in the nuclear rings of NGC~1097 and NGC~6951 (Barth et. al. 1995), and in 
the blue compact galaxy ESO-338-IG04 ({\"O}stlin et. al. 1998). 
  The magnitudes of YMCs reported in all these galaxies range from
$M_V \approx -10$ to $-15$, and the effective radii $R_e$ ($R_e$ = the
radius within which half of the light is contained) have been estimated to 
be of the order of a few parsec to about 20 pc, compatible with 
the objects deserving the designation ``young globular clusters''. 

  All of the systems mentioned above are relatively distant, but in fact
one does not have to go farther away than the Local Group in order to find
galaxies containing rather similar star clusters. The Magellanic Clouds have 
long been known to host star clusters of a type not seen in the Milky Way, 
i.e. compact clusters that are much more massive than Galactic open clusters 
(van den Bergh 1991, Richtler 1993), and in
many respects resemble globular clusters more than open clusters. Some of
the most conspicuous examples are the $10^7$ years old cluster 
in the centre of the 30 Doradus nebula in the LMC (Brandl et. al. 1996),
shining at an absolute visual magnitude of 
about -11, and the somewhat older object NGC~1866 (about $10^8$ years), also
in the LMC (Fischer et. al. 1992), which has an absolute
visual magnitude of $M_V \approx -9.0$. Even if these clusters are not
quite as spectacular as those found in genuine starburst galaxies, they
are still more massive than any of the open clusters seen in the Milky
Way today.  YMCs have been reported also in M33 (Christian \& Schommer 1988),
and in the giant Sc spiral M101 (Bresolin et. al. 1996).

  Taking into account the spread in the ages of the YMCs in the Antennae, 
Fritze - v. Alvensleben (1998) recovered a luminosity function (LF) resembling 
that of old globular clusters (GC's) to a very high degree when evolving the 
present LF to an age of 12 Gyr. Elmegreen \& Efremov (1997) point out the 
interesting fact that the {\it upper} end of the LF of old GC systems is 
very similar to that observed for YMCs, open clusters in the Milky Way, and 
even for HII regions in the Milky Way, and this is one of their arguments
in favour of the hypothesis that the basic mechanism behind the 
formation of all these objects is the same. They argue that massive clusters
are formed whenever there is a high pressure in the interstellar medium,
due to starbursts or other reasons as e.g. a high virial density 
(as in nuclear rings and dwarf galaxies). However, this doesn't seem to 
explain the presence of YMCs in apparently undisturbed disk galaxies like 
M33 and M101.

  So it remains a puzzling problem to understand why YMCs exist in 
certain galaxies, but not in others. In this paper we describe some first 
results from an investigation aiming at addressing this question.
It seems that YMC's can exist in a wide variety of host galaxy environments, 
and there are no clear systematics in the properties of the galaxies in 
which YMC's have been identified. And just like it is not clear how YMCs 
and old globular clusters are related to each other, one can also ask if 
the very luminous YMCs in mergers and starburst galaxies are basically the 
same type of objects as those in the Magellanic Clouds, M33 and M101.

\begin{table*}
\caption{The galaxies. In the first column each galaxy is identified by
its NGC number, the second column gives the morphological classification
taken from NED, right ascension and declination for equinox 2000.0 are in
columns 3 and 4. Apparent blue magnitude (from RC3) is in the 5th column,
the distance modulus is given in column 6, and the absolute blue magnitude
$M_B$ is in column 7.  The last column indicates which telescope was used 
for the observations 
(DK154 = Danish 1.54m. telescope, NOT = Nordic Optical Telescope).
The sources for the distances are as follows:
$^1$de Vaucouleurs 1963,
$^2$Carignan 1985,
$^3$Freedman et. al. 1992,
$^4$Bottinelli et. al. 1985,
$^5$Nearby Galaxies Catalog (Tully 1988)
$^6$de Vaucouleurs 1979a,
$^7$de Vaucouleurs 1979b,
$^8$Shanks 1997,
$^9$Karachentsev \& Drozdovsky 1998,
$^{10}$Freedman \& Madore 1988
$^{11}$Karachentsev et. al. 1996.
}
\label{tab:gal}
\begin{tabular}{llllllll}\hline
 Name    &  Type     & $\alpha$ (2000.0) & $\delta$ (2000.0) &
    $m_B$ & $m-M$   & $M_B$ & Obs. \\ \hline
NGC 45   & SA(s)dm   & 00:14:04     & $-23$:10:52      & 
    11.32 & $28.42\pm0.41^4$ & -17.13 & DK154 \\
NGC 247  & SAB(s)d   & 00:47:08     & $-20$:45:38      & 
    9.67 & $27.0\pm0.4^2$ & -17.40 & DK154 \\ 
NGC 300  & SA(s)d    & 00:54:53     & $-37$:41:00      & 
    8.72 & $26.66\pm0.10^3$ & -18.05 & DK154 \\
NGC 628  & SA(s)c    & 01:36:42     & +15:46:59      & 
      9.95  & $29.6\pm0.4^7$ & -19.77 & NOT \\
NGC 1156 & IB(s)m    & 02:59:43     & +25:14:15      & 
      12.32 & $29.46\pm0.15^{11}$ & -17.84 & NOT \\
NGC 1313 & SB(s)d    & 03:18:15     & $-66$:29:51      & 
    9.2  & $28.2^1$ & -19.03 & DK154 \\
NGC 1493 & SB(rs)cd  & 03:57:28     & $-46$:12:38      & 
      11.78 & $30.3^5$ & -18.62 & DK154 \\
NGC 2403 & SAB(s)cd  & 07:36:54     & +65:35:58      & 
      8.93  & $27.51\pm0.24^{10}$ & -18.73 & NOT \\
NGC 2835 & SAB(rs)c  & 09:17:53     & $-22$:21:20      & 
    11.01 & $28.93\pm0.42^4$ & -18.30 & DK154 \\
NGC 2997 & SA(s)c    & 09:45:39     & $-31$:11:25      & 
    10.06 & $29.9\pm0.4^7$ & -20.35 & DK154 \\
NGC 3184 & SAB(rs)cd & 10:18:17     & +41:25:27      & 
      10.36 & $29.5\pm0.4^7$ & -19.14 & NOT \\
NGC 3621 & SA(s)d    & 11:18:16     & $-32$:48:42      & 
      10.28 & $29.1\pm0.18^8$ & -19.21 & DK154 \\
NGC 4395 & SA(s)m    & 12:25:49     & +33:32:48      & 
      10.64 & $28.1^9$ & -17.47 & NOT \\
NGC 5204 & SA(s)m    & 13:29:36     & +58:25:04      & 
      11.73 & $28.4^{5}$ & -16.68 & NOT \\
NGC 5236 & SAB(s)c   & 13:37:00     & $-29$:51:58      & 
    8.20  & $27.84\pm0.15^6$ & -19.78 & DK154 \\
NGC 5585 & SAB(s)d   & 14:19:48     & +56:43:44      & 
      11.20 & $29.2^{5}$ & -18.00 & NOT \\
NGC 6744 & SAB(r)bc  & 19:09:45     & $-63$:51:22      & 
    9.14  & $28.5\pm0.4^7$ & -19.50 & DK154 \\
NGC 6946 & SAB(rs)cd & 20:34:52     & +60:09:14      & 
      9.61  & $28.7^{5}$ & -20.70 & NOT \\
NGC 7424 & SAB(rs)cd & 22:57:18     & $-41$:04:14      & 
    10.96 & $30.5^5$ & -19.54 & DK154 \\
NGC 7741 & SB(s)cd   & 23:43:53     & +26:04:35      & 
      11.84 & $30.8\pm0.4^{7}$ & -19.10 & NOT \\ 
NGC 7793 & SA(s)d    & 23:57:50     & $-32$:35:21      & 
    9.63  & $27.6\pm0.20^2$ & -18.04 & DK154 \\ \hline
\end{tabular}
\end{table*}

  We therefore decided to observe a number of nearby galaxies and look for 
populations of YMC's. The galaxies were mainly selected from the Carnegie 
Atlas (Sandage \& Bedke 1994), and in order to minimise the problems that 
could arise from extinction internally in the galaxies we selected galaxies 
that were more or less face-on. We tried to cover as wide a range in 
morphological properties as possible, although the requirement that the 
galaxies had to be nearby (because we would rely on ground-based observations) 
restricted the available selection substantially. The final sample consists 
of 21 galaxies out to a distance modulus of $m-M \approx 30$, for which 
basic data can be seen in Table \ref{tab:gal}. 

  In this paper we give an overview of our observations, and we discuss the 
main properties of the populations of YMCs in the galaxies in 
Table~\ref{tab:gal}.
In a subsequent paper (Larsen et. al. 1999) we will discuss
the correlations between the number of YMCs in a galaxy and various
properties of the host galaxies in more detail, and compare our data with 
data for starburst galaxies and mergers published in the literature.

\section{Observations and reductions}

  The observations were carried out partly with the Danish 1.54 m.
telescope and DFOSC (Danish Faint Object Spectrograph and Camera) at the 
European Southern Observatory (ESO) at La Silla, Chile, and partly with 
the 2.56 m. Nordic Optical Telescope (NOT) and ALFOSC (a DFOSC twin
instrument), situated at La Palma, Canary Islands. The data consists
of CCD images in the filters U,B,V,R,I and H$\alpha$. In the filters BVRI and 
H$\alpha$ we typically made 3 exposures of 5 minutes each, and 3 
exposures of 20 minutes each in the U band. Both the ALFOSC
and DFOSC were equipped with thinned, backside-illuminated 2 K$^2$ 
Loral-Lesser CCDs. The pixel scale in the ALFOSC is 0.189\arcsec /pixel and
the scale in the DFOSC is 0.40\arcsec /pixel, and the fields covered by these 
two instruments are $6.5\arcmin \times 6.5\arcmin$ and 
$13.7\arcmin \times 13.7\arcmin$, respectively.  All observations used in 
this paper were conducted under photometric conditions, with typical seeing 
values (measured on the CCD images) being 1.5\arcsec\ and 0.8\arcsec\ for 
the La Silla and La Palma data, respectively.

  During each observing run, photometric standard stars in the Landolt
(1992) fields were observed for calibration of the photometry. Some of
the Landolt fields were observed several times during the night at
different airmass in order to measure the atmospheric 
extinction coefficients. For the flatfielding we used skyflats exposed
to about half the dynamic range of the CCD, and in general each
flatfield used in the reductions was constructed as an average of
about 5 individual integrations.

  After bias subtraction and flatfielding, the three exposures in each 
filter were combined to a single image, and star clusters were identified 
using the {\bf daofind} task in DAOPHOT (Stetson 1987) on a 
background-subtracted \mbox{$V$-band} frame. Aperture photometry was then 
carried out with the DAOPHOT {\bf phot} task, using a small aperture 
radius (4 pixels for colours and 8 pixels for the \mbox{$V$-band} magnitudes) 
in order to minimise errors arising from the greatly varying background. 
Aperture corrections from the standard star photometry 
(aperture radius = 20 pixels) to the science data were
derived from a few isolated, bright stars in each frame.  Because the star 
clusters are not true point sources, no PSF photometry was attempted. A 
more detailed description of the data reduction procedure will be given 
in Larsen (1999).

  The photometry was corrected for Galactic foreground extinction using the
$A_B$ values given in the {\it Third Reference Catalogue of Bright Galaxies} 
(de Vaucouleurs et. al. (1991), hereafter RC3). 

\subsection{Photometric errors}
\label{sec:photerr}

The largest {\it formal} photometric errors as estimated by {\bf phot}
are those in the $U$ band, amounting to around 0.05 mag. for the 
faintest clusters. However, these error estimates are based on pure photon 
statistics and are not very realistic in a case like ours. Other 
contributions to the errors come from the standard transformation procedure, 
from a varying background, and from the fact that the clusters are not 
perfect point sources so that the aperture corrections become uncertain.

\begin{figure}
\epsfxsize=8cm
\epsfbox{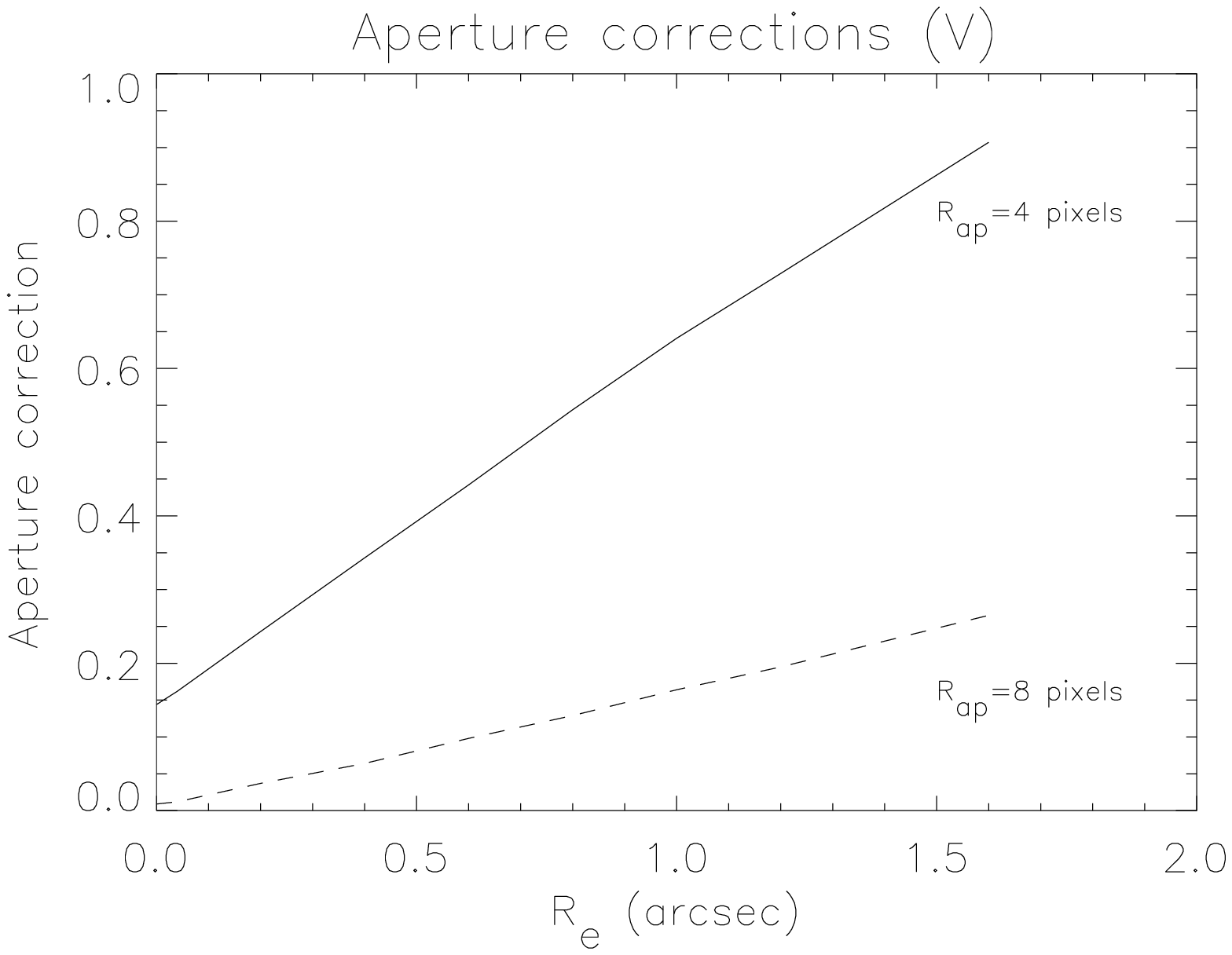}
\epsfxsize=8cm
\epsfbox{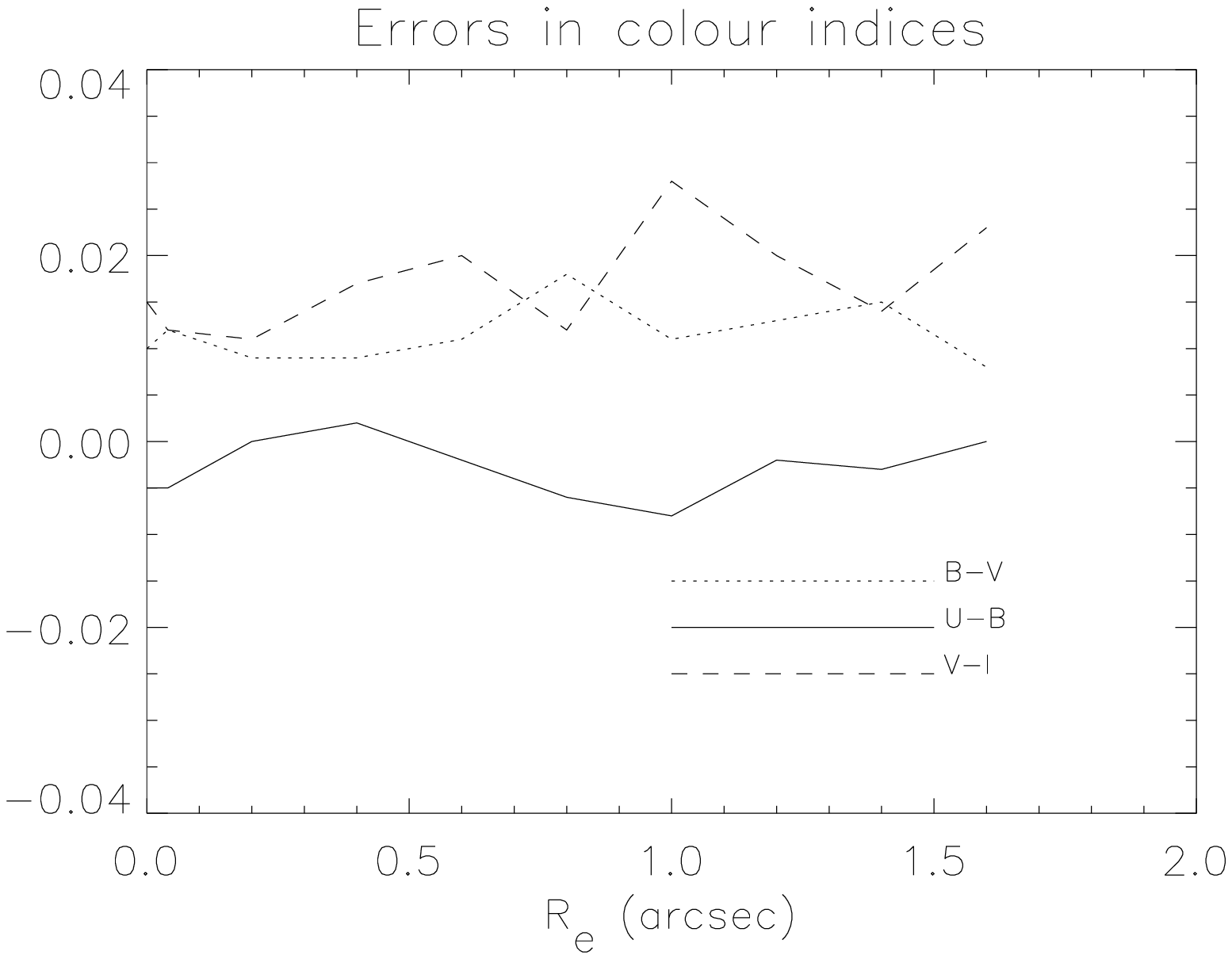}
\caption{
  Aperture corrections as a function of intrinsic cluster radius. 
{\it Top:} The errors in the \mbox{$V$-band} aperture corrections
for aperture radii of 4 and 8 pixels. {\it Bottom}: The corresponding 
errors in the colours for an aperture radius of 4 pixels.
\label{fig:apc}
}
\end{figure}

  The r.m.s. residuals of the standard transformations were between 
0.01 - 0.03 mags. in V, B-V and V-I, and between 0.04 and 0.06 mags. in U-B. 

  The errors in aperture corrections arising from the finite cluster
sizes were estimated by carrying out photometry on artificially generated 
clusters with effective radii in the range $R_e = 0 - 4$ pixels
(0\arcsec\ - 1.6\arcsec\ on the DFOSC frames). 1\arcsec\ corresponds to a 
linear distance of about 20 pc at the distance of typical galaxies in our 
sample, such as NGC~1313 and NGC~5236.  The artificial clusters were modeled 
by convolving the point-spread function (PSF) with MOFFAT15 profiles.

  The upper panel in Fig.~\ref{fig:apc} shows the errors in the 
aperture corrections for \mbox{$V$-band} photometry through aperture radii
$R_{ap} = 4$ pixels and $R_{ap} = 8$ pixels as a function of $R_e$, while
the lower panel shows the errors in the colour indices for $R_{ap} = 4$
pixels. At $R_e \approx 1\arcsec$, the error in \mbox{$V$-band} magnitudes
using $R_{ap} = 8$ pixels amounts to about 0.15 magnitudes. For a given
$R_e$, the errors in the {\it colours} are much smaller than the errors
in the individual bandpasses, so that accurate colours can be derived
through the small $R_e = 4$ pixels aperture without problems. This 
convenient fact has also been demonstrated by e.g. Holtzman et. al.  (1996).

\begin{figure}
\epsfxsize=85mm
\epsfbox{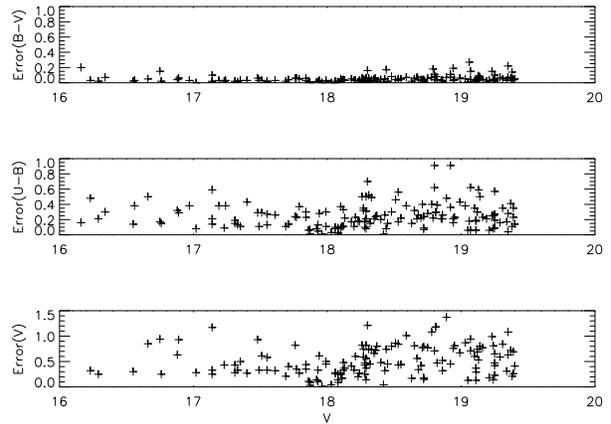}
\caption{The random errors in $V$, $B-V$ and $U-B$ 
as a function of $V$ magnitude for clusters in NGC~5236.
\label{fig:err_ran}
}
\end{figure}

  The random errors, primarily arising due to background fluctuations,
should in principle be evaluated individually for each cluster, since
they depend on the local environment of the cluster. Fig.~\ref{fig:err_ran}
shows the random errors for clusters in NGC~5236, estimated 
by adding artificial objects of similar brightness and colour near each 
cluster and remeasuring them using the same photometric procedure as
for the cluster photometry.  Again it is found that the errors in two 
different filters tend to cancel out when colour indices are formed.
The \mbox{$V$-band} errors are quite substantial, but we have chosen to
accept the large random errors associated with the use of an 
$R_{ap}=8$ pixels aperture in order to keep the effect of systematic
errors at a low level.

\section{Identification of star clusters}

\label{sec:id}

  After the photometry had been obtained, the first step in the analysis was 
to identify star cluster candidates, and to make sure that they were really 
star clusters and not some other type of objects. Possible sources of 
confusion could be compact HII regions, foreground stars, and individual 
luminous stars in the observed galaxies.  However, each of these objects can 
be eliminated by applying the following selection criteriae:

\begin{itemize}
  \item HII regions: These can be easily identified due to their $H\alpha$ 
	emission.
  \item Foreground stars: Because our galaxies are located at rather high 
        galactic latitudes, practically all foreground stars are redder than 
	\mbox{$B-V \approx 0.45$},
        whereas young massive star clusters will be bluer than this
        limit. Hence, by applying a $B-V$ limit of 0.45 we sort away
	the foreground stars while retaining the young massive cluster
	candidates. Remaining foreground stars could in many cases be
	distinguished by their position in two-colour diagrams, by
	their lack of angular extent, and by being positioned outside the
        galaxies.
  \item Individual luminous stars in the galaxies: We apply a brightness 
        limit of $M_V = -8.5$ for cluster candidates with $U-B > -0.4$ and
	$M_V = -9.5$ for candidates with $U-B < -0.4$. 
	The bluer objects are often found inside or near star forming
	regions, but the magnitude limit of $M_V = -9.5$ should prevent
	confusion with even very massive stars. 
\end{itemize}

  In addition to these selection criteriae it was found very useful to
generate colour-composite images using the I, U and H$\alpha$ exposures
and identify all the cluster candidates visually on these images. For
the ``red'' channel we used the H$\alpha$ exposures, for the ``green''
channel we used the I-band frames, and for the ``blue'' channel the
U-band frames.  In images
constructed like this, YMCs stand out very clearly as compact blue
objects, in contrast to HII regions which are distinctly red, and foreground
stars and background galaxies which appear green.

  Following the procedure outlined above, we ended up with a list of 
star cluster candidates in each galaxy.
The cluster nature of the detected objects was further verified by
examining their positions in two-colour diagrams (U-B,B-V and U-V,V-I),
and compare with model predictions for the colours of star clusters and
individual stars. In addition, we have been
able to obtain spectra of a few of the brightest star cluster candidates.
These will be discussed in a subsequent paper.

  The cluster samples may suffer from incompleteness effects.  In particular, 
we have deliberately excluded the youngest clusters which are still 
embedded in giant HII regions (corresponding to an age of
less than about $10^7$ years). Clusters which have 
intrinsic $B-V < 0.45$ will also slip out of the sample if their actual 
observed $B-V$ index is larger than 0.45 due to reddening internally in 
the host galaxy. 

\subsection{Counting clusters}
\label{sec:count}

The {\it specific frequency} for old globular cluster systems has
traditionally been defined as (Harris \& van den Bergh 1981):
\begin{equation}
  S_N = N_{\mbox{\scriptsize GC}} \times 10^{0.4 \times (M_V + 15)}
  \label{eq:sn}
\end{equation}
where $N_{\mbox{\scriptsize GC}}$ is the total number of globular clusters 
belonging to a galaxy 
of absolute visual magnitude $M_V$. Such a definition is a reasonable
way to characterise old globular cluster systems because 
$N_{\mbox{\scriptsize GC}}$ is a 
well-defined quantity, which can
be estimated with good accuracy due to the gaussian-like luminosity
function (LF) even if the faintest clusters are not directly observable. In
the case of young clusters it is more complicated to define a useful 
measure of the richness of the cluster systems, because the LF is no
longer gaussian and the number of young clusters that one finds in a galaxy 
depends critically on the magnitude limit applied in the survey. 
Nevertheless, we have defined a quantity equivalent to $S_N$ for the
{\it young} cluster systems:
\begin{equation}
  T_N = N_{\mbox{\scriptsize YMC}} \times 10^{0.4 \times (M_B + 15)}
  \label{eq:tn}
\end{equation}
$N_{\mbox{\scriptsize YMC}}$ is the number of clusters $N_B + N_R$ 
satisfying the criteriae described in Sec.~\ref{sec:id}. We have chosen to 
normalise $T_N$ to the \mbox{$B$-band} luminosity of the host galaxy because 
it can be looked up directly in the RC3 catalogue. 

\section{Results}

\subsection{Specific frequencies}

\begin{table}
\caption{Number of clusters identified in each of the galaxies.
$N_B$ refer to the 'blue' clusters (i.e. clusters with
$U-B < -0.4$ and $N_R$ refer to the 'red' clusters (clusters
with $U-B \ge -0.4$). The fourth column is the total number of
clusters, $N_B + N_R$. The quantities $T_N$ and $T_{N,C}$ are defined in 
Sect.~\ref{sec:count}. $^1$Only the central parts of the galaxies
were covered by our observations.
\label{tab:tn}
}
\begin{tabular}{lllllp{15mm}}\hline
Name  & $N_B$ & $N_R$ & $N_{\mbox{\scriptsize YMC}}$ & $T_N$ & $T_{N,C}$ \\ 
\hline
NGC 45   &  1 &  2 &  3 & $0.42\pm 0.29$ & - \\
NGC 247  &  1 &  2 &  3 & $0.32\pm 0.23$ & - \\
NGC 300  &  1$^1$ &  2$^1$ &  3$^1$ & 0.18$\pm 0.11^1$ & - \\
NGC 628  & 27 & 12 & 39 & $0.48\pm 0.19$ & $0.50\pm0.20$
                                           $0.57\pm0.23$ \\
NGC 1156 & 13 &  9 & 22 & $1.61\pm 0.41$ & $1.81\pm 0.46$
                                           $3.09\pm 0.92$ \\
NGC 1313 & 17 & 29 & 46 & $1.12\pm 0.27$ & $1.36\pm0.32$
                                           $1.63\pm0.39$ \\
NGC 1493 &  0 &  0 &  0 &    0 & - \\
NGC 2403 &  4$^1$ & 10$^1$ & 14$^1$ & $0.45\pm0.16^1$ & \\
NGC 2835 &  7 &  2 &  9 & $0.43\pm 0.22$ & - \\
NGC 2997 & 20 & 14 & 34 & $0.25\pm 0.10$ & $0.28\pm 0.11$
                                           $0.31\pm 0.12$ \\
NGC 3184 &  3 & 10 & 13 & $0.28\pm 0.13$ & - \\
NGC 3621 & 22 & 23 & 45 & $0.93\pm 0.21$ & $1.14\pm 0.26$
                                           $1.40\pm 0.32$ \\
NGC 4395 &  0 &  2 &  2 & $0.21\pm 0.15$ & - \\
NGC 5204 &  0 &  7 &  7 & $1.49\pm 0.63$ & - \\
NGC 5236 & 55 & 96 & 151 & $1.75\pm 0.28$ & $1.98\pm 0.32$
                                            $3.29\pm 0.58$ \\
NGC 5585 &  1 &  8 &  9 & $0.57\pm 0.22$ & - \\
NGC 6744 & 12 &  6 & 18 & $0.43\pm 0.22$ & \\ 
NGC 6946 & 76 & 31 &107 & $0.56\pm 0.11$ & $0.66\pm0.14$
					   $0.81\pm0.17$ \\
NGC 7424 &  7 &  3 & 10 & $0.15\pm 0.09$ & - \\
NGC 7741 &  0 &  0 &  0 &    0 & - \\
NGC 7793 & 12 &  8 & 20 & $1.21\pm 0.35$ & $1.33\pm0.39$
					   $1.62\pm0.47$ \\
(LMC)    &  1 &  7 &  8 & $0.57$ & - \\
(M33)    &  0 &  1 &  1 & $0.04$ & - \\ \hline
\end{tabular}
\end{table}

\begin{figure}
\epsfxsize=9cm
\epsfbox{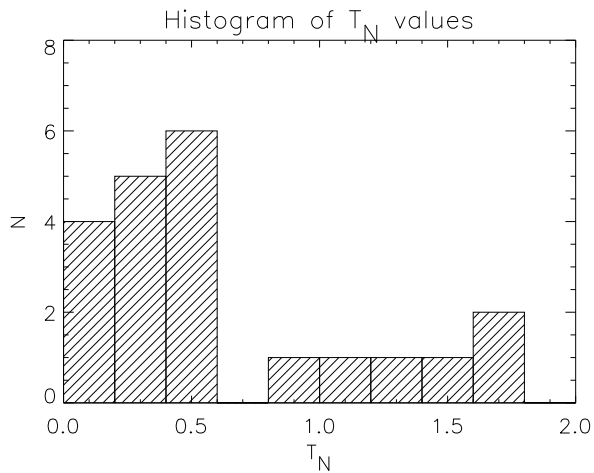}
\caption{Histogram of (uncorrected) $T_N$ values.
\label{fig:tnhisto}
}
\end{figure}

\begin{figure*}
\begin{minipage}{8.5cm}
\epsfxsize=8.5cm
\epsfbox{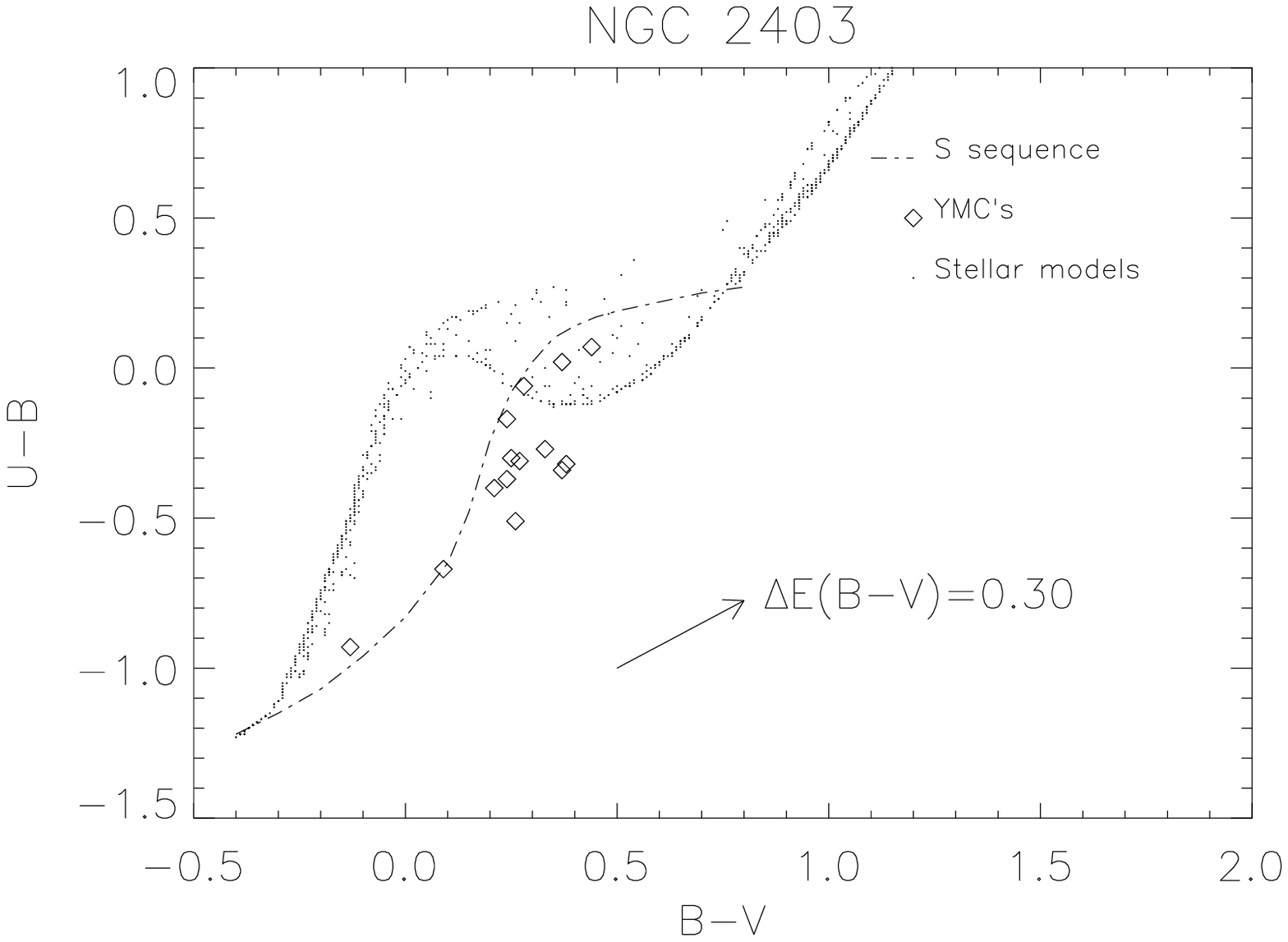}
\end{minipage}
\begin{minipage}{8.5cm}
\epsfxsize=8.5cm
\epsfbox{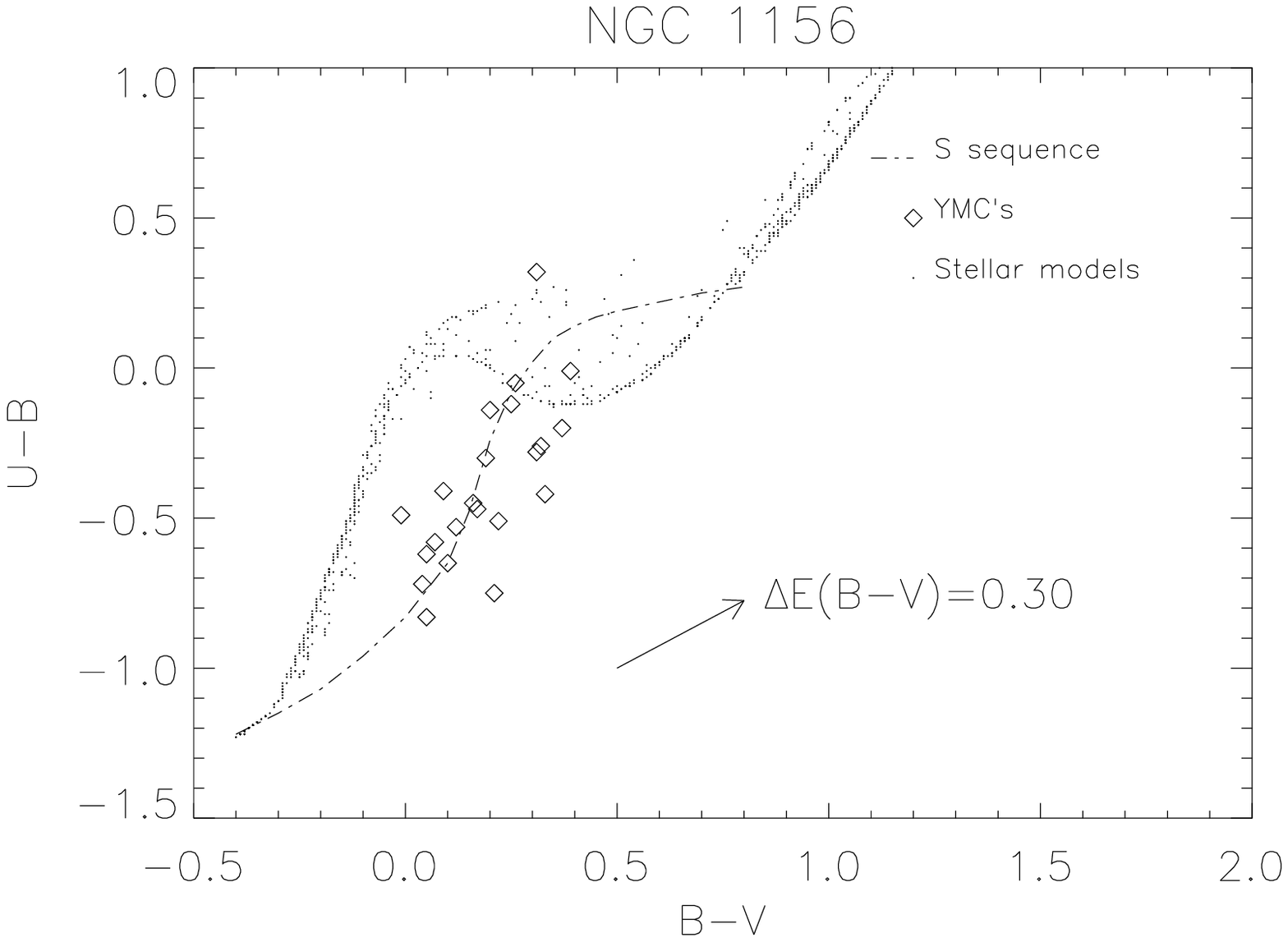}
\end{minipage}
\\
\begin{minipage}{8.5cm}
\epsfxsize=8.5cm
\epsfbox{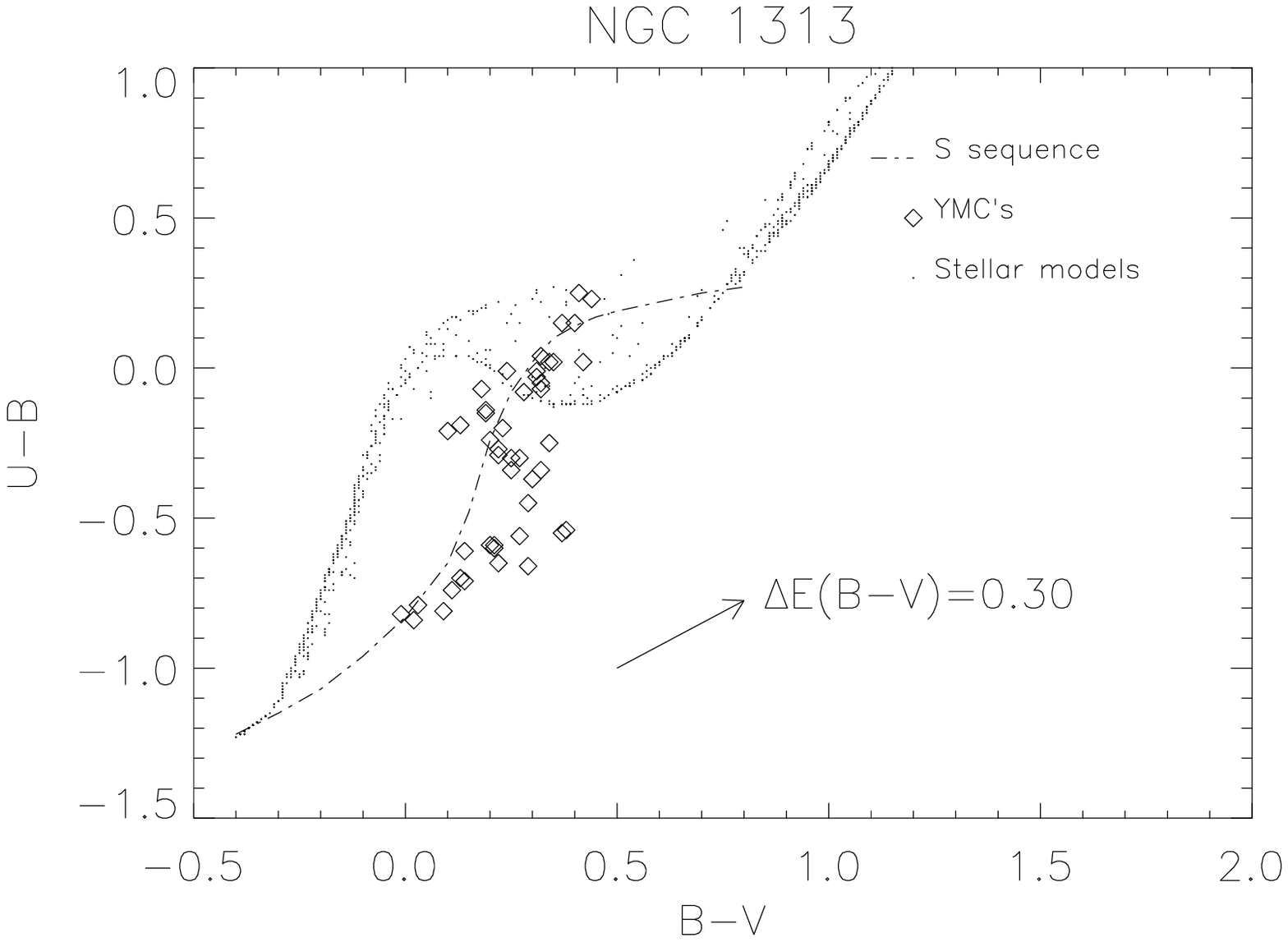}
\end{minipage}
\begin{minipage}{8.5cm}
\epsfxsize=8.5cm
\epsfbox{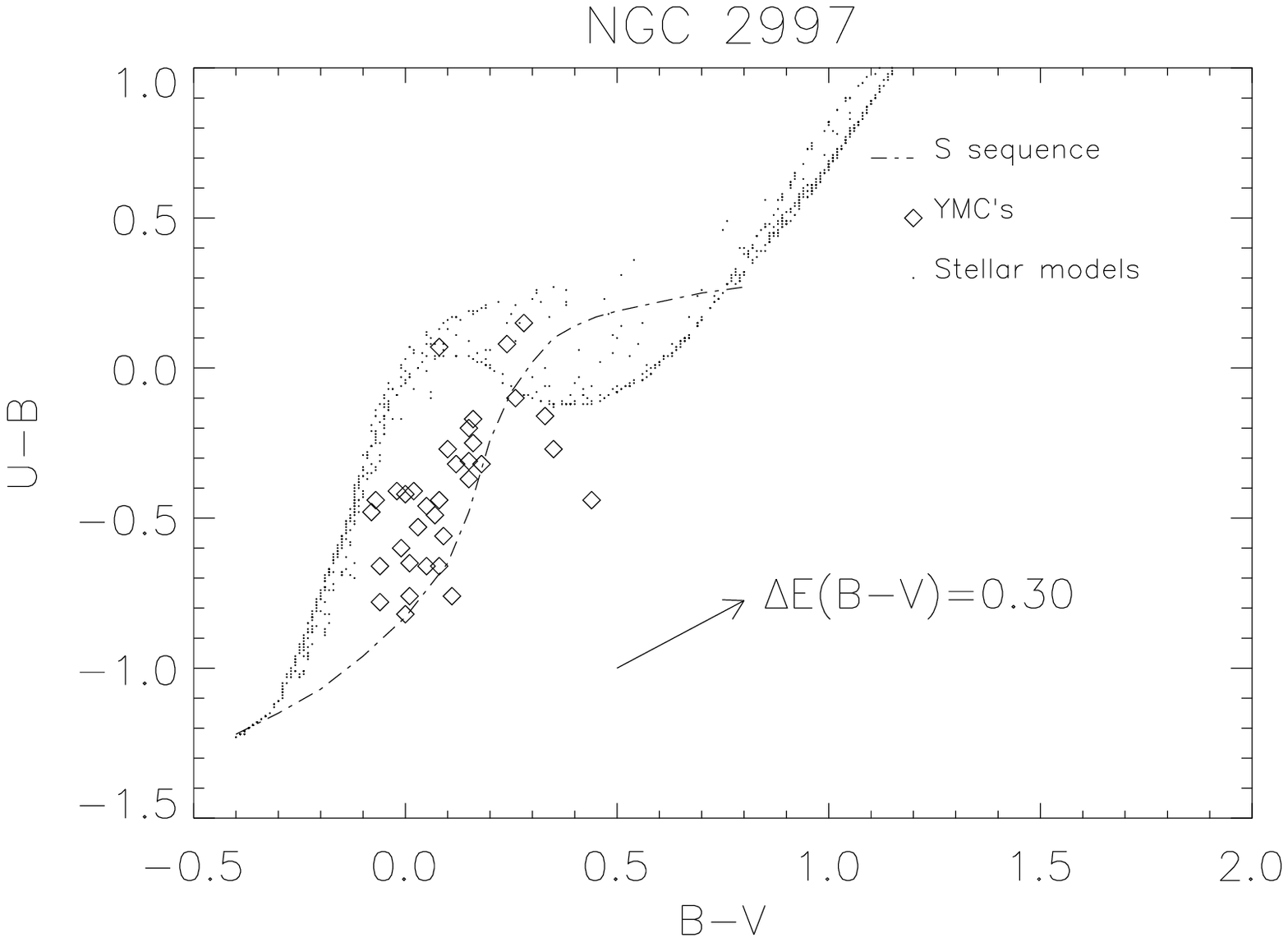}
\end{minipage}
\\
\begin{minipage}{8.5cm}
\epsfxsize=8.5cm
\epsfbox{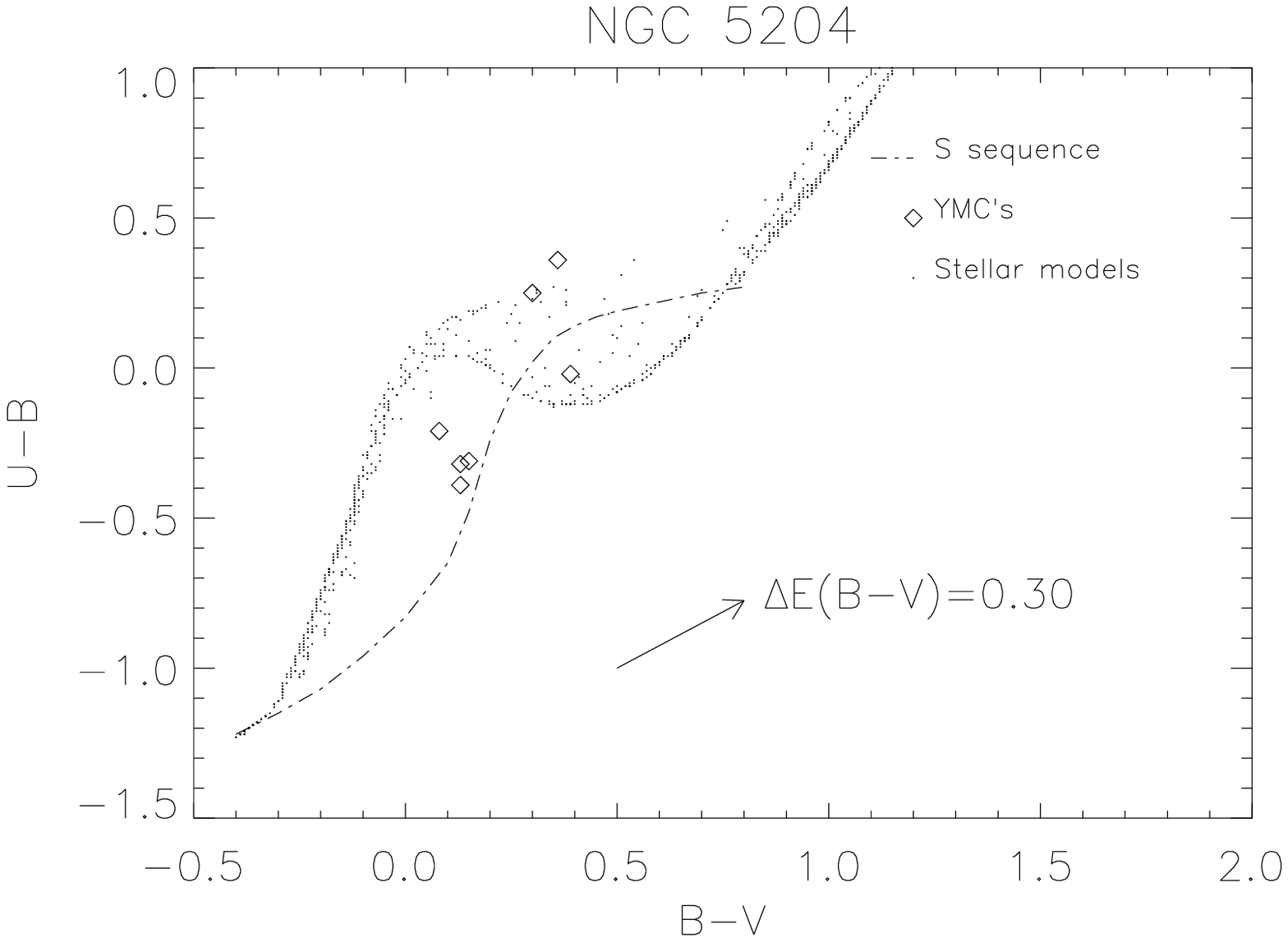}
\end{minipage}
\begin{minipage}{8.5cm}
\epsfxsize=8.5cm
\epsfbox{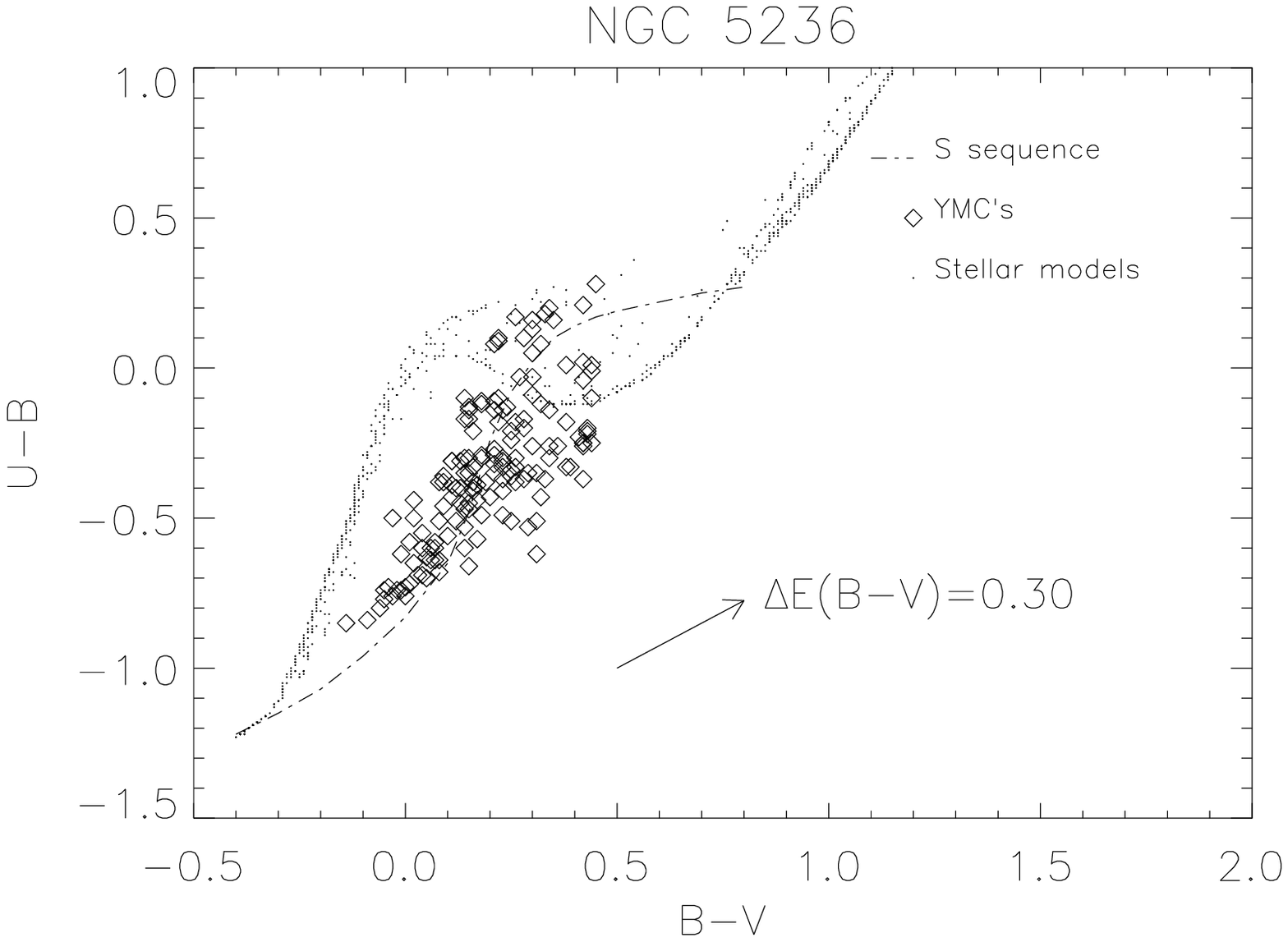}
\end{minipage}
\caption{U-B vs. B-V diagrams for the clusters in six galaxies, 
compared to stellar models by Bertelli et. al. 1994 and evolutionary
synthesis models of stellar clusters.
\label{fig:bv_ub}
}
\end{figure*}

\begin{figure}
\epsfxsize=9cm
\epsfbox{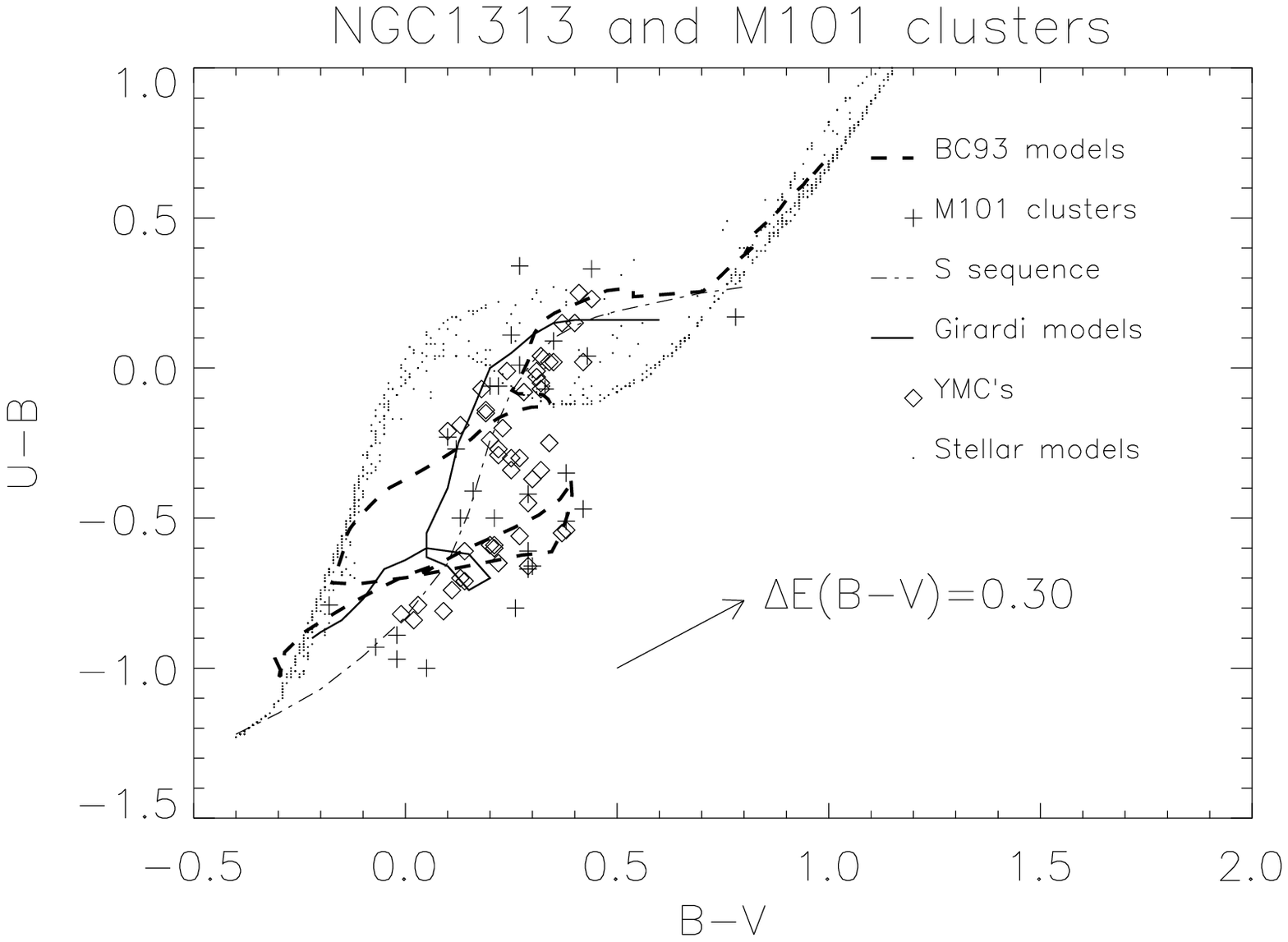}
\caption{Comparison between photometry of clusters in NGC~1313 and M101.
'+' markers indicate the HST photometry of clusters in M101 by Bresolin
et. al. 1996.
\label{fig:m101}
}
\end{figure}

  In Table \ref{tab:tn} we give the number of clusters identified in
each of the observed galaxies. The columns labeled $N_B$ and $N_R$ 
refer to the number of ``blue'' and ``red'' clusters respectively, according
to the definition that ``blue'' clusters are clusters with $U-B < -0.4$
(and hence $M_V < -9.5$) whereas the ``red'' clusters have $U-B \ge -0.4$
(and $M_V < -8.5$). See also Sect.~\ref{sec:id}. The data for the
LMC are from Bica et. al. (1996) and those for M33 are from Christian \&
Schommer (1988).

  The ``specific frequencies'' $T_N$ for the galaxies in our sample
are given in the fourth column of Table \ref{tab:tn}. The number 
$N_{\mbox{\scriptsize YMC}} = N_B + N_R$ used to derive $T_N$ is the total 
number of clusters, ``red'' and ``blue'', detected in each galaxy.

  The errors on $T_N$ were estimated taking into consideration only the 
uncertainties of the absolute magnitudes of the host galaxies resulting 
from the distance errors as given in Table~\ref{tab:gal} and poisson 
statistics of the cluster counts. However, it is clear that this is not a 
realistic estimate of the total uncertainties of the $T_N$ values. Another 
source of uncertainty arises from incompleteness effects, particularly for 
the more distant galaxies. For all galaxies with more than 20 clusters we 
estimated the incompleteness by adding artificial clusters with magnitudes 
of 18.0, 18.5 $\ldots$ 21.0, and testing how many of the artificially 
added clusters were detected by DAOFIND in {\it all} of the filters 
$U$,$B$ and $V$. Because the completeness depends critically on
the size of the objects, we carried out completeness tests for artificial
clusters with $R_e = 0$ pc and $R_e = 20$ pc in each galaxy. The numbers
of clusters actually detected in each of the magnitude bins [18.25 - 18.75],
[18.75 - 19.25] $\ldots$ [20.75 - 21.25] were then corrected by the fraction 
of artificial clusters recovered in the corresponding bin, and finally
the ``corrected'' $T_N$ values were derived. These are given in the
last column of Table~\ref{tab:tn}, labeled $T_{N,C}$, for point sources
(first line) and objects with $R_e = 20$ pc (second line). See
Larsen~(1999) for more details on the completeness corrections.

  One additional source of errors affecting $T_N$ which remains
uncorrected, is the fact that an uncertainty in the distance also affects
the magnitude limit for detection of star clusters . If a galaxy is more 
distant (or nearby) than the value we have adopted, our limit corresponds to 
a ``too bright'' (too faint) absolute magnitude, and we have underestimated 
(overestimated) the number of clusters. Hence, the true $T_N$ errors are 
somewhat larger than those given in Table~\ref{tab:tn}, but they depend
on the cluster luminosity function. If the clusters
follow a luminosity function of the form $\phi(L)dL \propto L^{-1.78}dL$
(Whitmore \& Schweizer 1995) then a difference in the magnitude limit
of $\Delta M_V = 0.1$ would lead to a difference in the cluster counts of 
about 7\%.

A histogram of the uncorrected $T_N$ values (Fig.~\ref{fig:tnhisto})
shows that a wide range of $T_N$ values are present within our sample. Many 
of the galaxies in the lowest bins contain only a few massive clusters or none
at all, but a few galaxies have much higher $T_N$ values than the average. 
The most extreme $T_N$ values are found in NGC~1156, NGC~1313, NGC~3621, 
NGC~5204 and NGC~5236. The galaxy NGC~2997 also hosts a very rich cluster 
system, but the $T_N$ value is probably severely underestimated because of 
the large distance of NGC~2997 which introduces significant incompleteness 
problems.  A similar remark applies to two other distant galaxies observed at 
the Danish 1.54 m. telescope, NGC~1493 and NGC~7424, while all the 
remaining galaxies in Table~\ref{tab:gal} are either more nearby, or have been
observed at the NOT in better seeing conditions, and hence their $T_N$
values are believed to be more realistic.

\subsection{Two-colour diagrams}

In Fig.~\ref{fig:bv_ub} we show the $B-V, U-B$ diagrams for 
six cluster-rich galaxies. These plots also include the so-called ``S'' 
sequence defined by Girardi et. al. (1995, see also Elson \& Fall (1985)), 
represented as a dashed line. The ``S'' sequence is essentially an age 
sequence, derived as a fit to the average colours of bright LMC clusters in
the $U-B, B-V$ diagram. The age increases as one moves along the 
S-sequence from blue to red colours.  The colours of 
our cluster candidates are very much compatible
with those of the S-sequence, especially if one considers that there
is a considerable scatter around the S-sequence also for Magellanic Cloud
clusters (Girardi et. al. 1995). Also included in the diagrams are stellar 
models by Bertelli et. al. (1994) (dots), in order to demonstrate that the 
position of clusters within such a diagram is distinctly different from that 
of single stars. Already from Fig.~\ref{fig:bv_ub} one can see that there 
is a considerable age spread among the clusters in each galaxy, with the red
cut-off being due to our selection criteriae. The reddening vector
corresponding to a reddening of $E(B-V) = 0.30$ is shown in each plot
as an arrow, and it is quite clear that the spread along the S-sequence
cannot be entirely due to reddening effects.

  Bresolin et. al. (1996) used HST data to carry out photometry for star
clusters in the giant Sc-type spiral M101. A comparison between their
data and photometry for clusters in one of our galaxies (NGC~1313) is shown 
in Fig.~\ref{fig:m101}. It is evident that the colours of clusters 
in the two galaxies are very similar. NGC~1313 was chosen as an illustrative
example because it contains a relatively rich cluster system, although
not so rich that the diagram becomes too crowded.

  In Fig.~\ref{fig:m101} we have also included a curve 
showing the colours of star clusters according to the population synthesis 
models of Bruzual \& Charlot (1993, hereafter BC93). The agreement between 
the synthetic and observed colours
is very good for \mbox{U-B $>-0.3$}, but for \mbox{U-B $<-0.3$} the B-V 
colours of the BC93 models are systematically too blue compared to our data
and the S sequence.  The ``red loop'' that extends out to 
\mbox{B-V$\approx$0.3} and \mbox{U-B$\approx$-0.5} is due to the appearance 
of red supergiants at an age of about $10^7$ years (Girardi \& Bica 1993) 
and is strongly metallicity dependent.
  Girardi et. al. (1995) constructed population synthesis models based on a 
set of isochrones by Bertelli et. al. (1994) and found very good agreement
between the S-sequence and their synthetic colours. The models (solar
metallicity) are included in Fig.~\ref{fig:m101} as a solid line.  In these 
models the ``red loop'' is not as pronounced as in the BC93 models, and the 
youngest models are in general not as blue as those of BC93, resulting
in a much better fit to the observed cluster colours. 

\subsection{Ages and masses} 
\label{sec:age_mass}

\begin{figure*}
\begin{minipage}{8.5cm}
\epsfxsize=8.5cm
\epsfbox{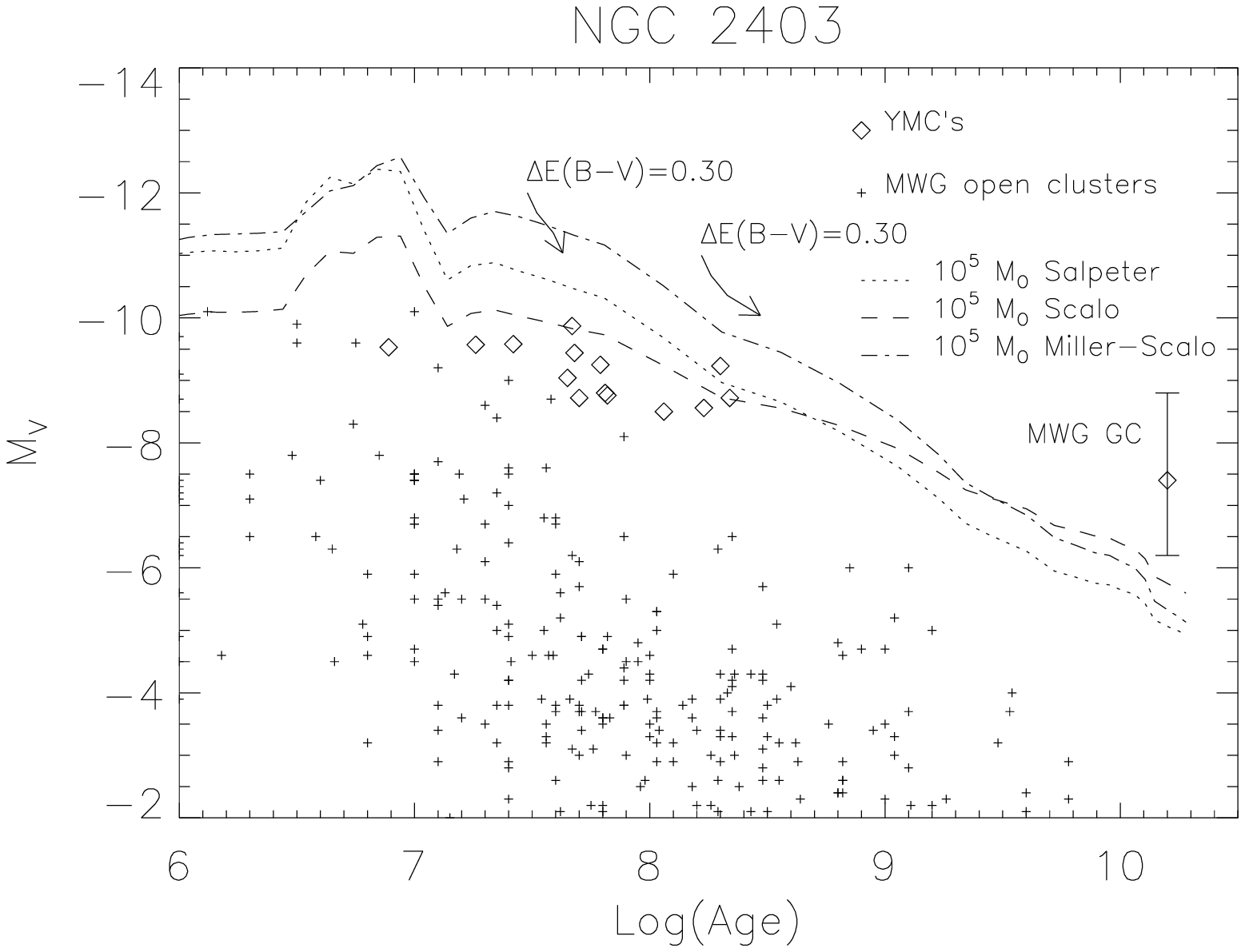}
\end{minipage}
\begin{minipage}{8.5cm}
\epsfxsize=8.5cm
\epsfbox{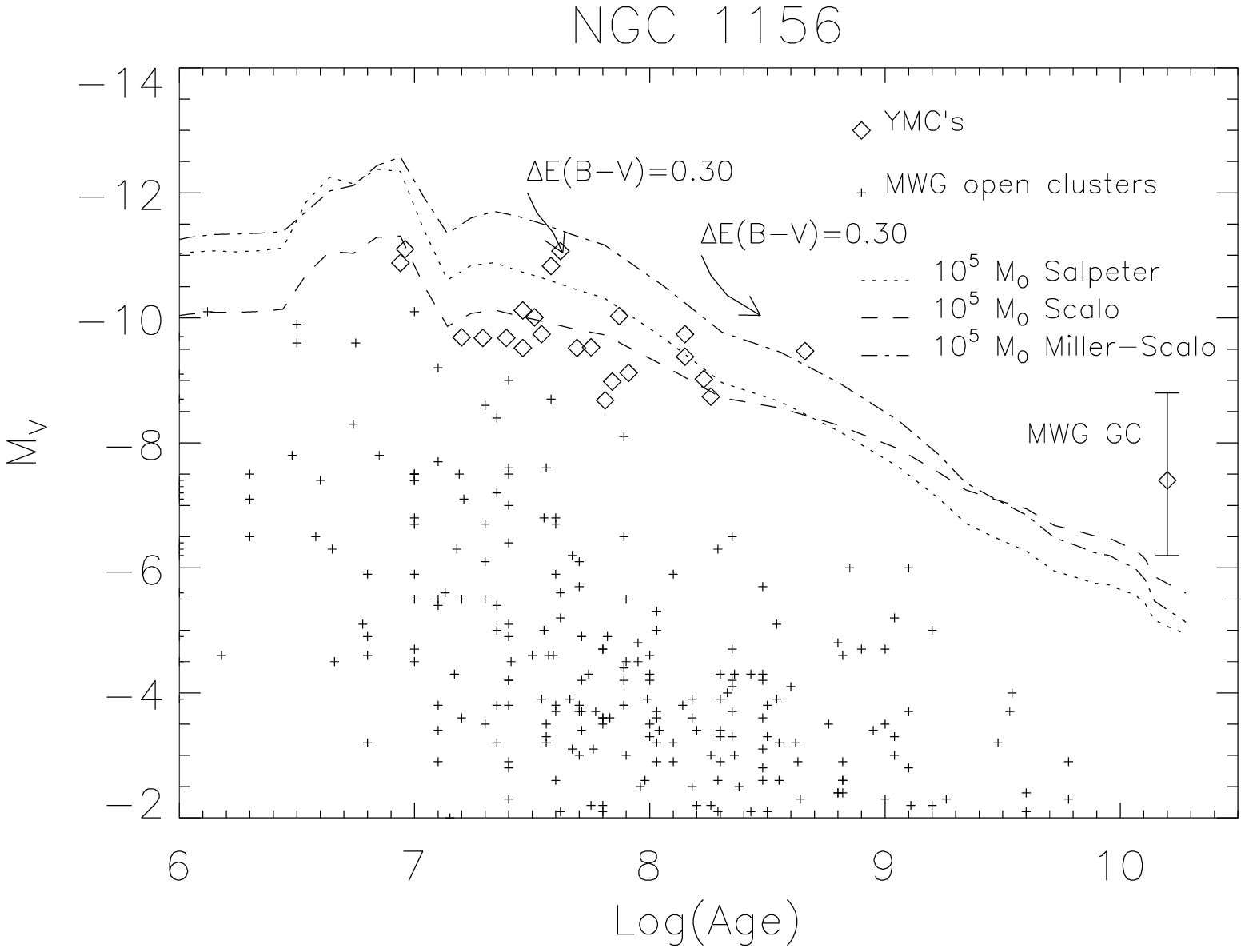}
\end{minipage}
\\
\begin{minipage}{8.5cm}
\epsfxsize=8.5cm
\epsfbox{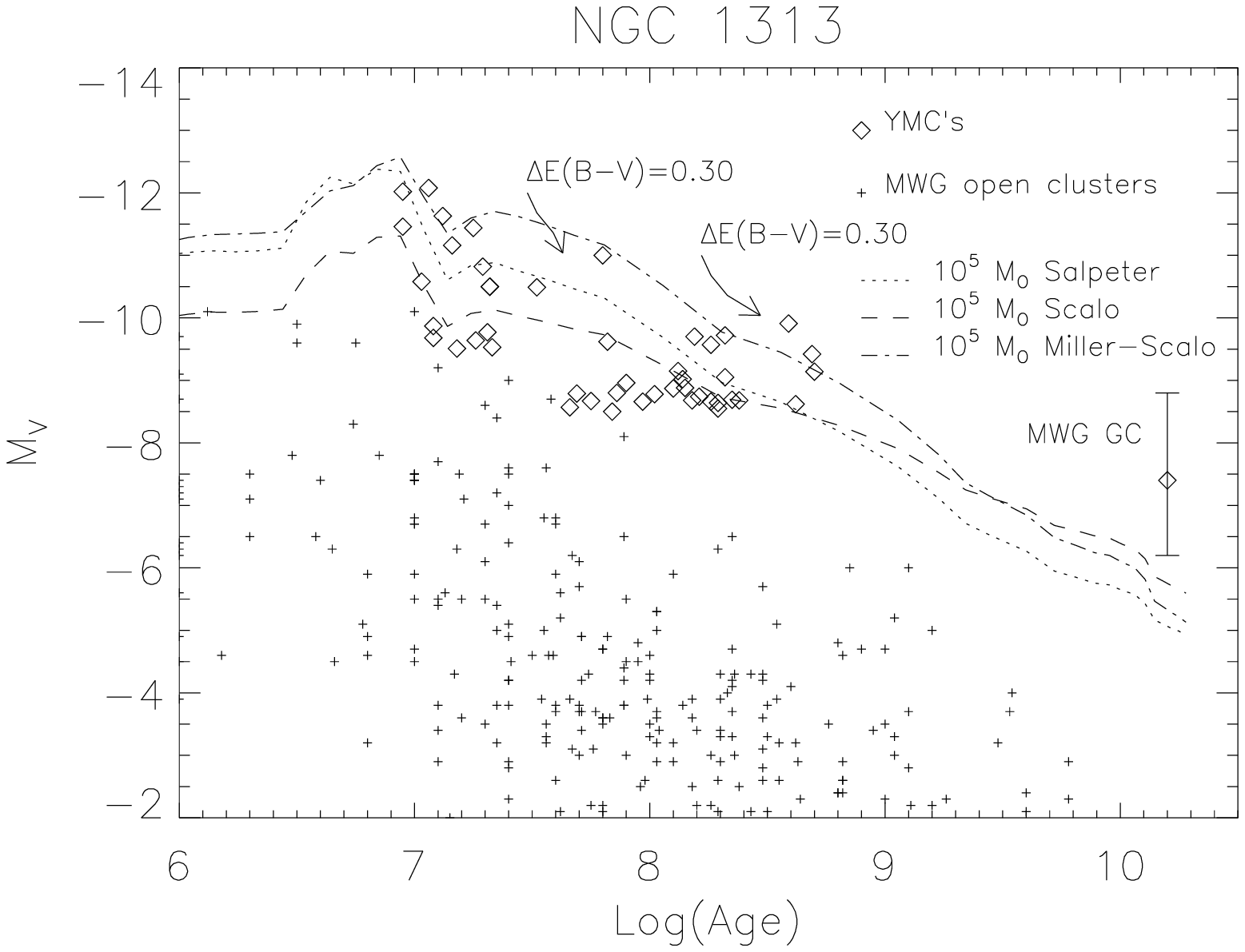}
\end{minipage}
\begin{minipage}{8.5cm}
\epsfxsize=8.5cm
\epsfbox{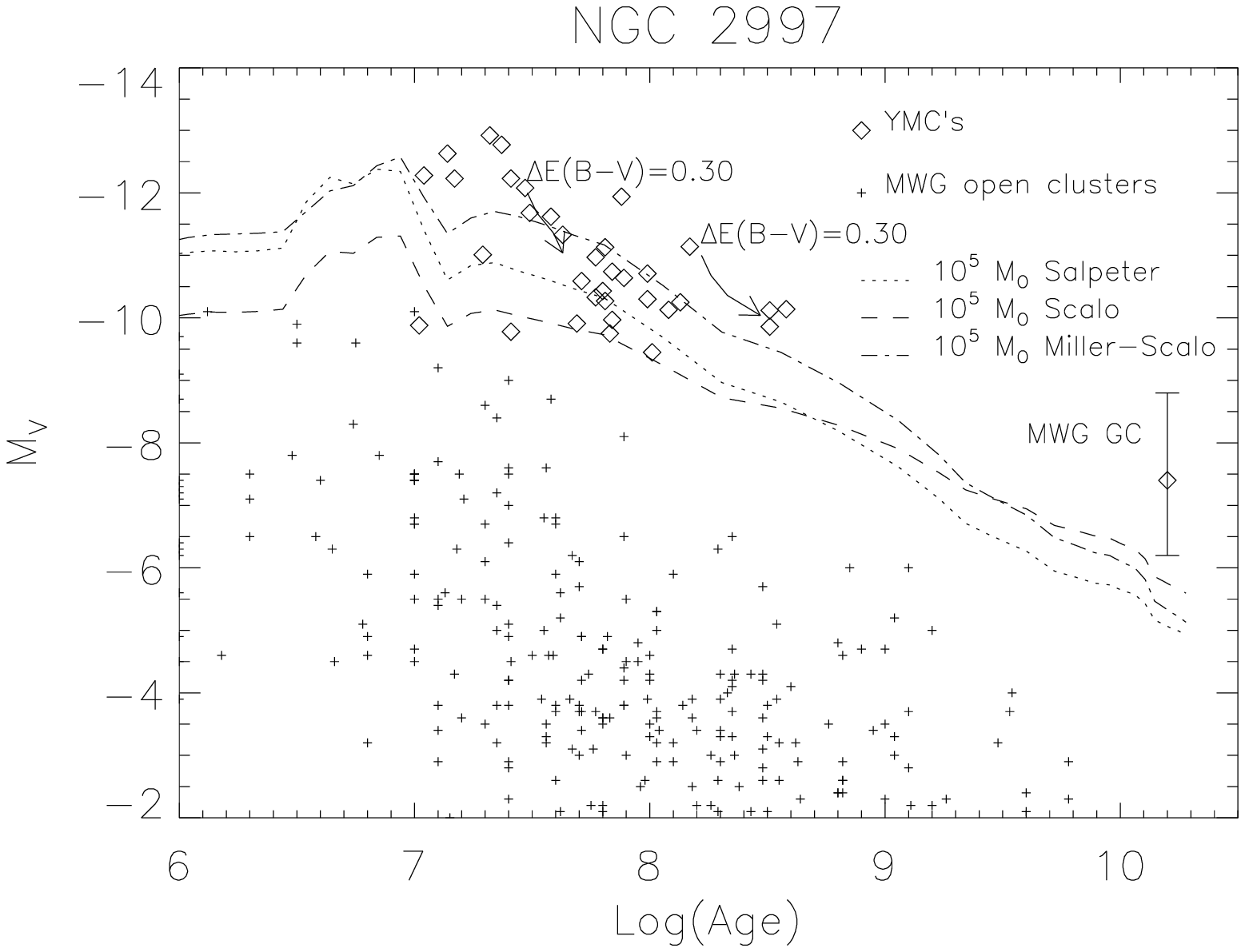}
\end{minipage}
\\
\begin{minipage}{8.5cm}
\epsfxsize=8.5cm
\epsfbox{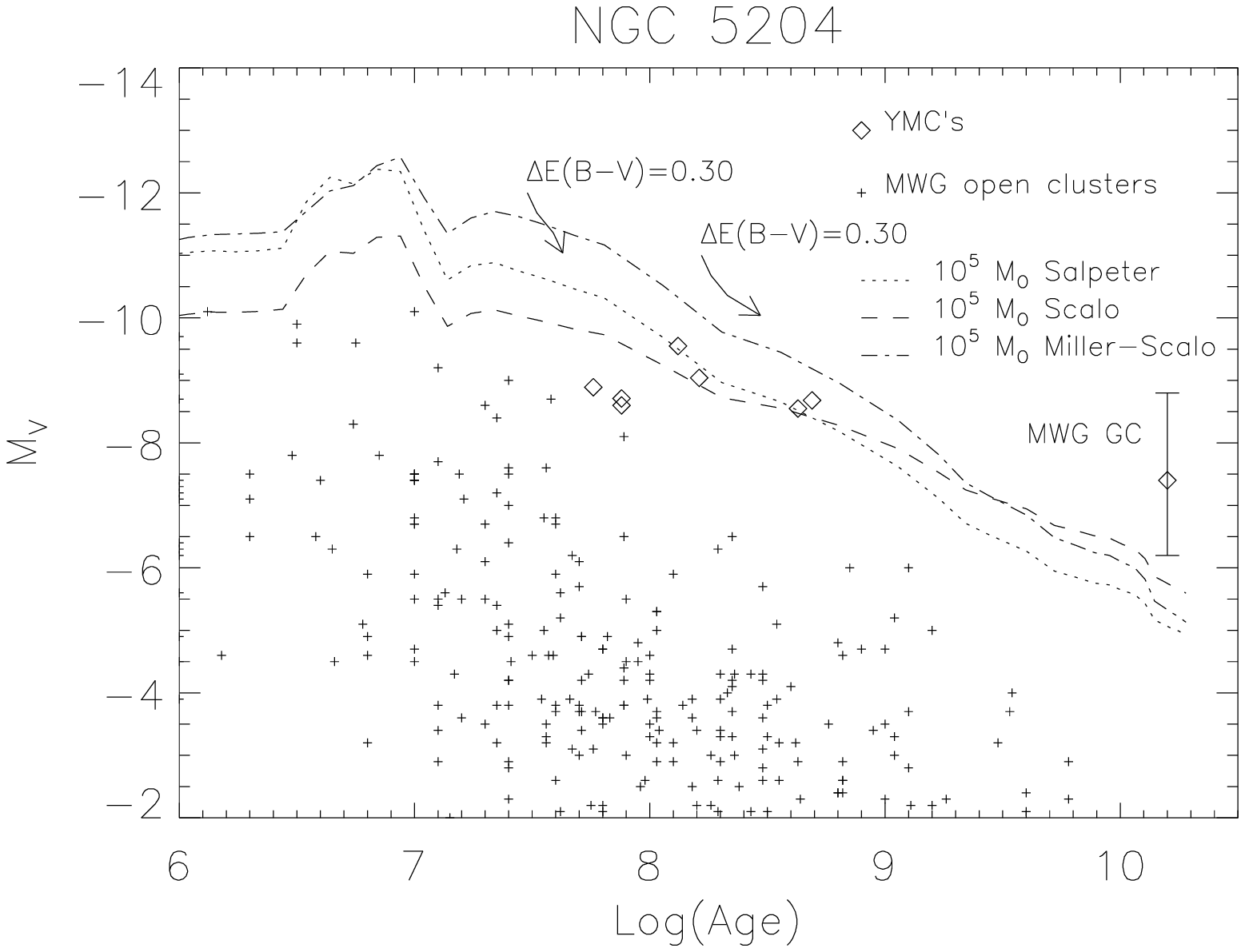}
\end{minipage}
\begin{minipage}{8.5cm}
\epsfxsize=8.5cm
\epsfbox{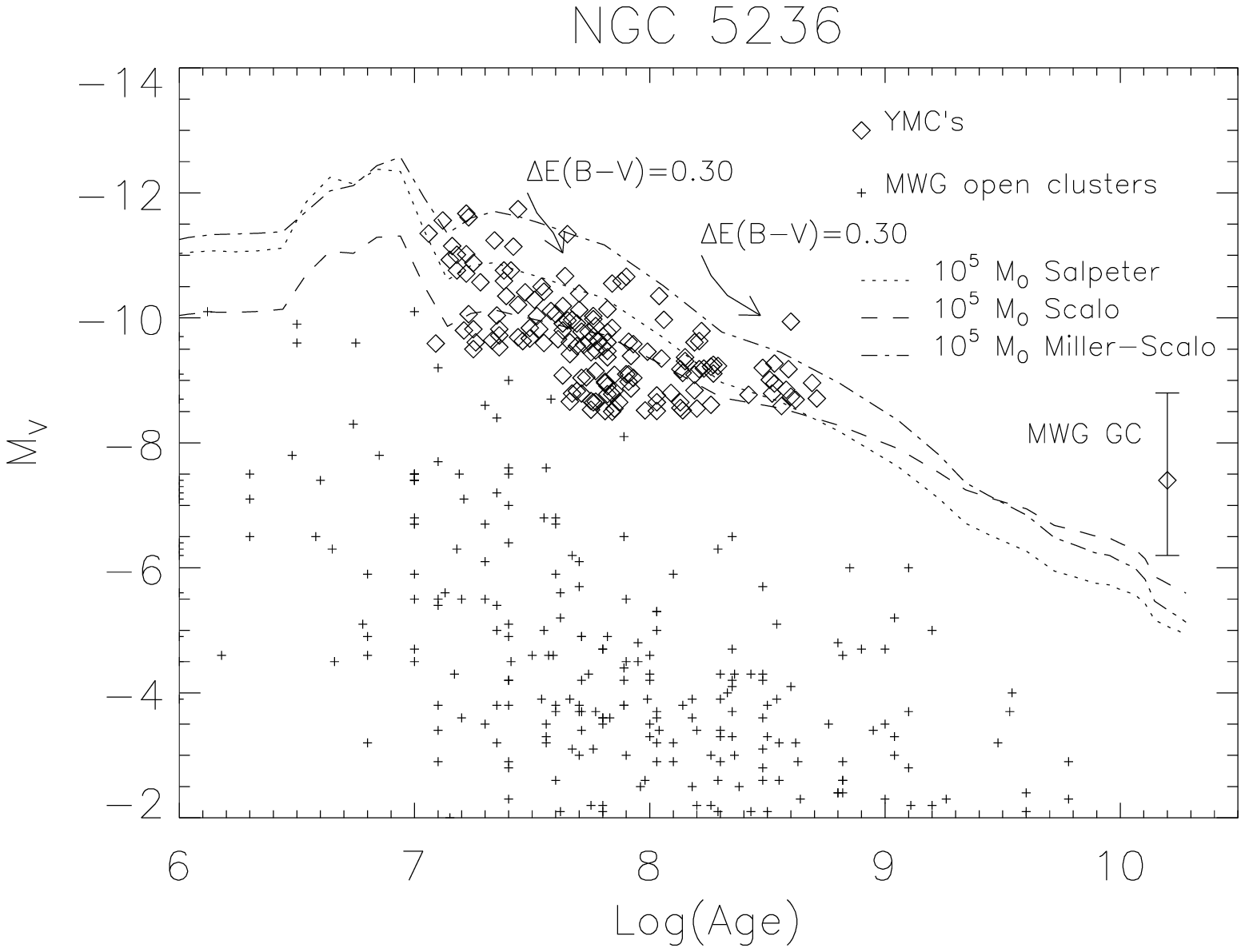}
\end{minipage}
\caption{Absolute visual magnitudes versus ages for the clusters in
six galaxies, compared with data for open clusters in the Milky Way 
(Lyng{\aa} 1982). Population synthesis models by Bruzual \& Charlot (1993)
for three different IMFs, corresponding to a mass of $10^5 M_{\odot}$,
have been included in each plot.
\label{fig:age_mv}
}
\end{figure*}

  A direct determination of the mass of an unresolved star cluster 
requires a knowledge of the M/L ratio, which in turn depends on many other
quantities, in particular the age and the IMF of the cluster. However,
if one assumes that the IMF does not vary too much from one star cluster to
another, then the luminosities alone should facilitate a comparison of 
star clusters {\it with similar ages}. 

  Applying the S-sequence age calibration to the clusters in our sample, 
the luminosities of each cluster can then be directly compared to Milky Way 
clusters of similar age, as shown in Fig.~\ref{fig:age_mv}. Ages and absolute 
visual magnitudes for Milky Way open clusters are from the 
Lyng{\aa} (1982) catalogue, and are represented in each plot as small crosses.
In the diagrams in Fig.~\ref{fig:age_mv} we have 
also indicated the effect of a reddening of E(B-V) = 0.30. In these plots 
the ``reddening vector'' depends in principle on the original position
of the cluster within the (U-B,B-V) diagram from which the age was derived,
but we have included two typical reddening vectors, corresponding to two
different ages.

  In all of the galaxies in Fig.~\ref{fig:age_mv} but NGC~2403, the absolute 
visual magnitudes of the brightest clusters are 2 - 3 magnitudes brighter 
than the upper limit of Milky Way open clusters of similar ages. Accordingly 
they should be nearly 10 times more massive. In the case of NGC~2403,
the most massive clusters are not significantly more massive than open 
clusters found in the Milky Way.  Fig.~\ref{fig:age_mv} also confirms the 
suspicion that the cluster data in NGC~2997 are incomplete, particularly for 
$M_V > -10$.

  We have included population synthesis models for the luminosity evolution
of single-burst stellar populations of solar metallicity by BC93 in 
Fig.~\ref{fig:age_mv}, scaled to a total 
mass of $10^5 M_{\odot}$. Models for three different IMFs are plotted: 
Salpeter~(1955), Miller-Scalo~(1979) and Scalo~(1986), all covering a
mass range from 0.1 - 65 $M_{\odot}$. 
The different assumptions about the shape of the IMF obviously affect the
evolution of the $M_V$ magnitude per unit mass quite strongly, 
and unfortunately the effect is most severe just in the age interval we
are interested in. The difference between the Miller-Scalo and the
Scalo IMF amounts to almost 2 magnitudes, but in any case the most massive
clusters appear to have masses around $10^5 M_{\odot}$.

  In Fig.~\ref{fig:age_mv} we have also indicated the location of a 
``typical''  old globular cluster system with an error bar centered on 
the coordinates 15 Gyr, $M_V = -7.4$ and with $\sigma_V = 1.2$ mags. Although 
the comparison of masses at high and low age based on population synthesis 
models is extremely sensitive to the exact shape of the IMF, it seems that 
the masses of the young massive star clusters are at least within the range 
of ``true'' globular clusters.  

  Reddening effects alone are unlikely to affect the derived ages to a 
high degree, as a scatter along the ``reddening vectors'' in 
Fig.~\ref{fig:age_mv} would then be expected. Basically this would 
mean that one would expect a much steeper rate of decrease in $M_V$ vs. the 
derived age, while the observed relation between age and the upper luminosity 
limit is in fact remarkably compatible with that predicted by the models.
The comparison with model calculations implies that the upper mass 
limit for clusters must have remained relatively unchanged over the entire 
period during which clusters have been formed in each galaxy.

\section{Notes on individual galaxies}
\label{sec:notes}

\subsection{NGC 1156}

  This is a Magellanic-type irregular galaxy, currently undergoing an
episode of intense star formation. Ho et. al. 1995 noted that the 
spectrum of NGC~1156 resembles that of the ``W-R galaxy'' NGC~4214.
NGC~1156 is a completely isolated galaxy, so the starburst could not have 
been triggered by interaction with other galaxies. We have found a 
number of massive star clusters in NGC~1156.

\subsection{NGC 1313}

This is an SB(s)d galaxy of absolute $B$ magnitude $M_B = -18.9$.
de Vaucouleurs 1963 found a distance modulus of $m-M = 28.2$, which we adopt.  
The morphology of NGC~1313
is peculiar in the sense that many detached sections of star formation are
found, particularly in the south-western part of the galaxy. There is
also a ``loop'' extending about 1.5 Kpc (projected) to the east of the bar 
with a number of HII regions and massive star clusters.  Another 
interesting feature is that one can see an extended, elongated diffuse 
envelope of optical light,
with the major axis rotated $45^{\circ}$ relative to the central bar 
of NGC~1313, embedding the whole galaxy. It has been suggested by 
Ryder et. al. (1995) that the diffuse 
envelope surrounding NGC~1313 is associated with galactic cirrus known to 
exist in this part of the sky (Wang \& Yu 1995), but this explanation
does not seem likely since it would require a very perfect alignment
of the centre of NGC~1313 with the diffuse light. Also, the outer boundary
of the active star-forming parts of galaxy coincide quite well with the
borders of the more luminous parts of the envelope.
In our opinion the most likely explanation is that the
diffuse envelope is indeed physically associated with NGC~1313 itself.

  Walsh \& Roy (1997) determined O/H abundances for 33 HII regions in 
NGC~1313, and found no radial gradient. This makes NGC~1313 the most
massive known barred spiral without any radial abundance gradient.

  NGC~1313 hosts a rich population of massive star clusters. 
When looking at the plot in Fig.~\ref{fig:age_mv} it seems that there is
a concentration of clusters at log(Age) $\approx$ 8.3 or roughly
200 Myr. We emphasize that this should be confirmed by a more thorough 
study of the cluster population in this galaxy, and in particular it would be
very useful to be able to detect fainter clusters in order to improve the
statistics. If this is real it could imply that some kind of
event stimulated the formation of massive star clusters in NGC~1313
a few hundred Myr ago, perhaps the accretion of a companion galaxy.
A second ``burst'' of cluster formation seems to have been taking place
very recently, and is maybe going on even today.

\subsection{NGC 2403}
 
  NGC~2403 is a nearby spiral, morphologically very similar to M33 apart 
from the fact that NGC~2403 lacks a distinct nucleus. It is a textbook
example of an Sc-type spiral, and it is very well resolved on our NOT
images. A photographic survey of star clusters
in NGC~2403 was already carried out by Battistini et. al. 1984, who
succeeded in finding a few YMC candidates. NGC~2403 spans more than
20 $\times$ 20 arcminutes in the sky, so we have been able
to cover only the central parts using the ALFOSC. Within the central
$6.5\arcmin \times 6.5\arcmin$ (about 6$\times$6 Kpc) we have located 14 
clusters altogether, 
but the real number of clusters in NGC~2403 should be significantly higher, 
taking into account the large fraction of the galaxy that we haven't covered, 
and considering the fact that in the other galaxies we have studied, many 
clusters are located at considerable distances from the centre.

\subsection{NGC 2997}

  NGC~2997 is an example of a ``hot spot'' galaxy (Meaburn et. al. 1982) with
a number of UV luminous knots near the centre. Walsh et. al. (1986)
studied the knots and concluded that they are in fact very massive
star clusters, and Maoz et. al. (1996) further investigated the central
region of NGC~2997 using the HST. On an image taken with the repaired HST
through the F606W filter they identified 155 compact sources, all with
diameters of a few pc. Of 24 clusters detected in the F606W filter as well
as in an earlier F220W image, all have colours implying ages less
than 100 Myr and masses $\ge 10^4 M_{\odot}$. Maoz et. al. (1996) conclude
that the clusters in the centre of NGC~2997 will eventually evolve into
objects resembling globular clusters as we know them in the Milky Way
today.

  In our study we have found a number of massive star clusters also
outside the centre of NGC~2997. Taking the numbers at face value, the cluster 
system does not appear to be as rich as that of NGC~5236, but with
better and more complete data we would expect to see a number of YMCs in 
NGC~2997 that could rival that in NGC~5236.

\subsection{NGC 3621}

  This galaxy is at first sight a quite ordinary late-type spiral, and has
not received much attention. It was observed with the HST by Rawson et. al.
(1997) as part of the {\it Extragalactic Distance Scale Key Project}, and
cepheids were discovered and used to derive a distance modulus of 29.1.
  
  Our data show that NGC~3621 contains a surprisingly high number of massive 
star clusters. The galaxy is rather inclined ($i=51^\circ$, Rawson et. al.
1997), and nearly all the clusters are seen projected on the near side of 
the galaxy, so a number of clusters on the far side may be hidden from our 
view. Ryder \& Dopita (1993) noted a lack of HII regions on 
the far side of the galaxy, and pointed out that there is also a quite
prominent spiral arm on the near side that doesn't appear to have a
counterpart on the far side. So it remains possible that the excess of
young clusters and HII regions on the near side is real. 

\subsection{NGC 5204}

  NGC~5204 is a companion to the giant Sc spiral M101. The structure of the
HI in this galaxy is that of a strongly warped disk (Sicotte \& Carignan
1997), and one could speculate that this is related to tidal interaction
effects with M101. Sicotte \& Carignan (1997) also find that the dark matter
halo of NGC~5204 contributes significantly to the mass even in the inner
parts.

The high $T_N$ value of this galaxy is a consequence of its very low $M_B$ 
rather than a high absolute number of clusters - we found only 7 clusters 
in this galaxy. Curiously, all of the 7 clusters belong to the ``red'' class,
suggesting that no new clusters are being formed in NGC~5204 at the moment. 

\subsection{NGC 5236}

  NGC 5236 (M83) is a grand-design barred spiral of type SBc, striking by 
its regularity and its very high surface brightness - the highest among 
the galaxies in our sample. The absolute visual magnitude is
$M_V = -20.0$ (de Vaucouleurs et. al. 1983).  NGC~5236 is currently 
undergoing a burst of star formation in the nucleus as well as in the spiral 
arms. 

  A study in the rocket UV (Bohlin et. al. 1990) has already revealed the 
presence of a number of very young massive star clusters inside the HII 
regions of NGC~5236, and HST observations of the nucleus (Heap et. al. 1993)
showed an arc of numerous OB clusters near the centre of the galaxy.
These clusters were found to have absolute visual magnitudes in the range
from $M_V = -10.4$ to $M_V = -13.4$. and typical radii of the order of
4 pc. Masses were estimated to be between $10^4$ and $10^5$ $M_{\odot}$.

Our investigation adds a large number of massive star clusters in NGC~5236 
also outside the centre and the HII regions. In terms of absolute numbers
the cluster system of NGC~5236 is by far the richest in our sample, and
in particular there is a large number of clusters in the ``red'' group.
This may be partly due to reddening effects although Fig.~\ref{fig:age_mv}
shows that there is in fact a large intrinsic age spread among the
clusters in NGC~5236. 

\subsection{NGC 6946}

\begin{figure}
\epsfxsize=8cm
\epsfbox{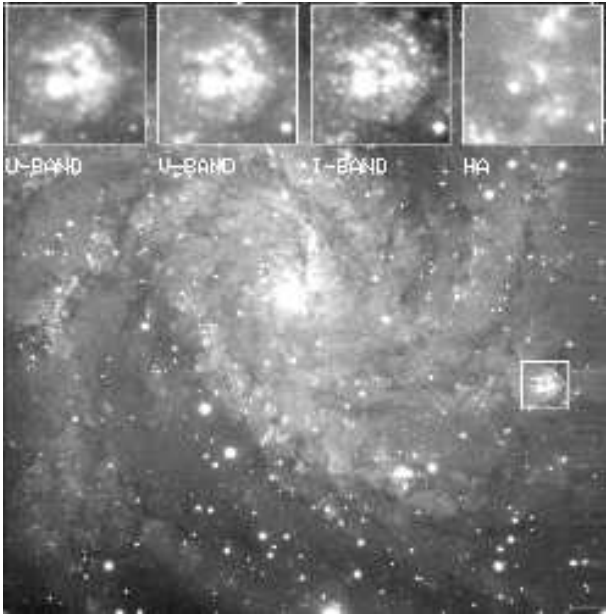}
\caption{A V-band image of NGC 6946. The cluster discussed in the text is
the luminous object located to the lower left of the centre of the 
bubble-like structure.
\label{fig:n6946}
}
\end{figure}

  The study of NGC~6946 is complicated by the fact that it is located at
low galactic latitude ($b = 12^{\circ}$), and there is an interstellar 
absorption of
$A_B = 1.6$ magnitudes and a large number of field stars towards this
galaxy. NGC~6946 is nevertheless a well-studied galaxy, and we 
also chose to include it in our sample, reasoning that star clusters 
should be recognizable as extended objects on the NOT data. 

  The chemical abundances of HII regions in NGC~6946 were studied by
Ferguson et. al. (1998), who concluded that their data were consistent
with a single log-linear dependence on the radius. At 1.5-2 optical radii 
(defined by the B-band 25th magnitude isophote) they measured abundances 
of O/H of about 10\%-15\% of the solar value, and N/O of about 20\% - 25\%
of the solar value.

  Among the approximately 100 clusters we have identified in NGC~6946, one 
stands out as particularly striking (Fig.~\ref{fig:n6946}). This cluster is 
apparently a very young object, located in one of the spiral arms at a 
distance of 4.4 Kpc from the centre, and with an impressive visual luminosity 
of $M_V = -13$. Using a deconvolution-like algorithm (Larsen 1999), the 
effective radius was estimated to be about 15 pc. The cluster is located 
within a bubble-like structure with a diameter of about 550 pc, containing 
numerous bright stars and perhaps some less massive clusters. On optical 
images this structure is very conspicuous, but it is not visible on the mid-IR 
ISOCAM maps by Malhotra et. al. (1996). There are no traces of $H\alpha$ 
emission either, except for a small patch at the very centre of the 
structure.

\subsection{LMC and M33}

  For these galaxies, we have adopted data from the literature.

  As mentioned in the introduction, both the LMC and M33 contain
young star clusters that are more massive than the open clusters seen in
the Milky Way.  However, as is evident from Table \ref{tab:tn}, only one 
cluster in M33 is a YMC according to our criteriae. The LMC, on the other 
hand, contains a relatively rich cluster population, with 7 clusters in the 
``red'' group and 1 cluster in the ``blue'' group.  The cluster R136 in 
the 30 Doradus nebula of the LMC has not been included in the data for
Table \ref{tab:tn} because of its location within a giant HII region.
Compared to the other galaxies in our sample, the LMC ranks among the
relatively cluster-rich ones, but it is also clear that a cluster population
like the one of the LMC is by no means unusual.

  Because the LMC is so nearby, the limiting magnitude for detection of
clusters is obviously much fainter than in the other galaxies in our 
sample, and the Bica et. al. (1996) catalogue should certainly be complete
down to our limit of $M_V = -8.5$, corresponding to $V = 10.25$ (taking
into account an absorption of about 0.25 mags. towards the LMC). If the LMC
was located at the distance of most of the galaxies in our sample we 
would probably not have detected 8 clusters, but a somewhat smaller number,
and the $T_N$ value would have been correspondingly lower. This should be 
kept in mind when comparing the data for the LMC with data for the rest of 
the galaxies in the sample.

\section{Radial density profiles of cluster systems}

\begin{figure*}
\begin{minipage}{8.5cm}
\epsfxsize=8.5cm
\epsfbox{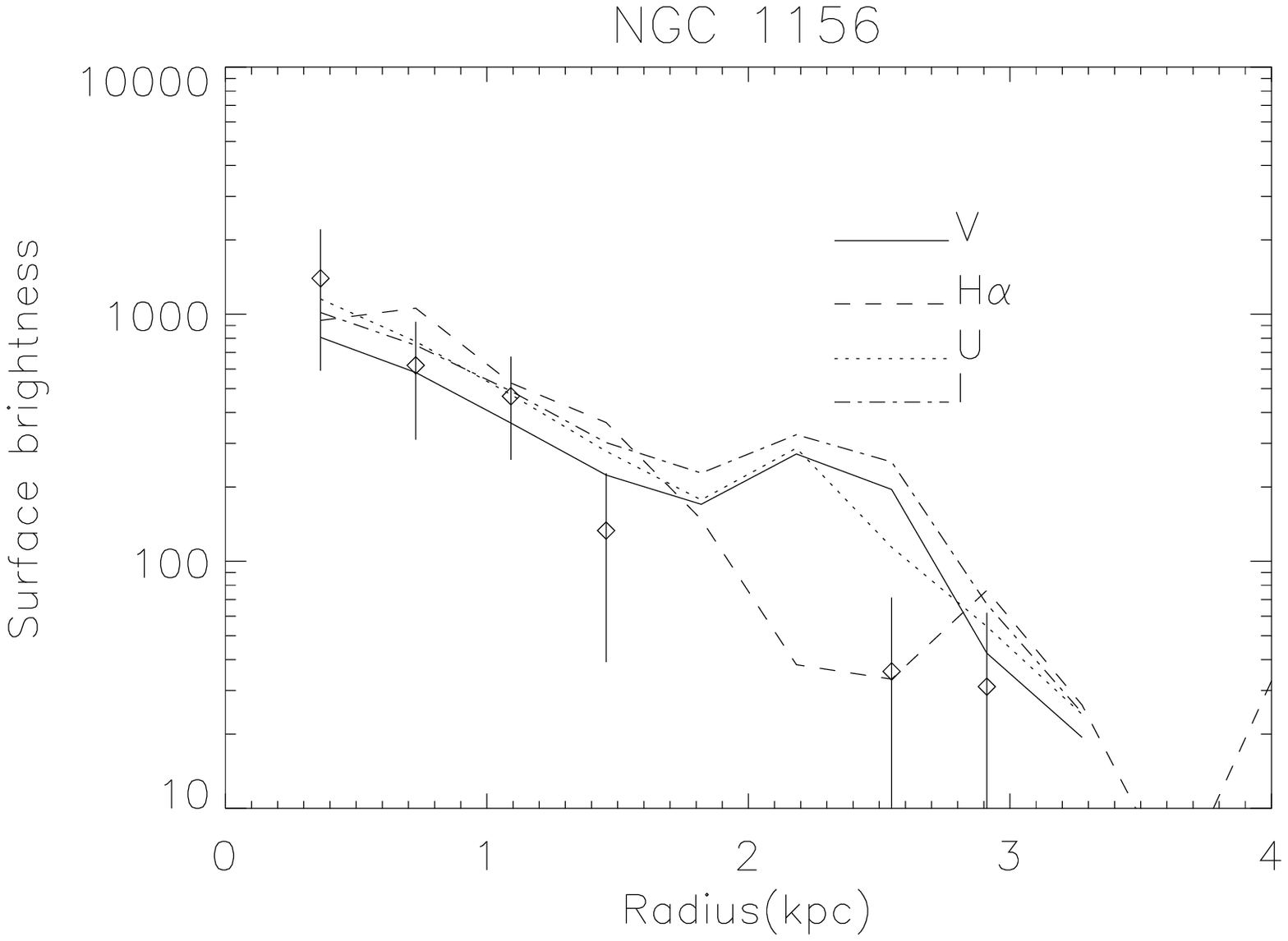}
\end{minipage}
\begin{minipage}{8.5cm}
\epsfxsize=8.5cm
\epsfbox{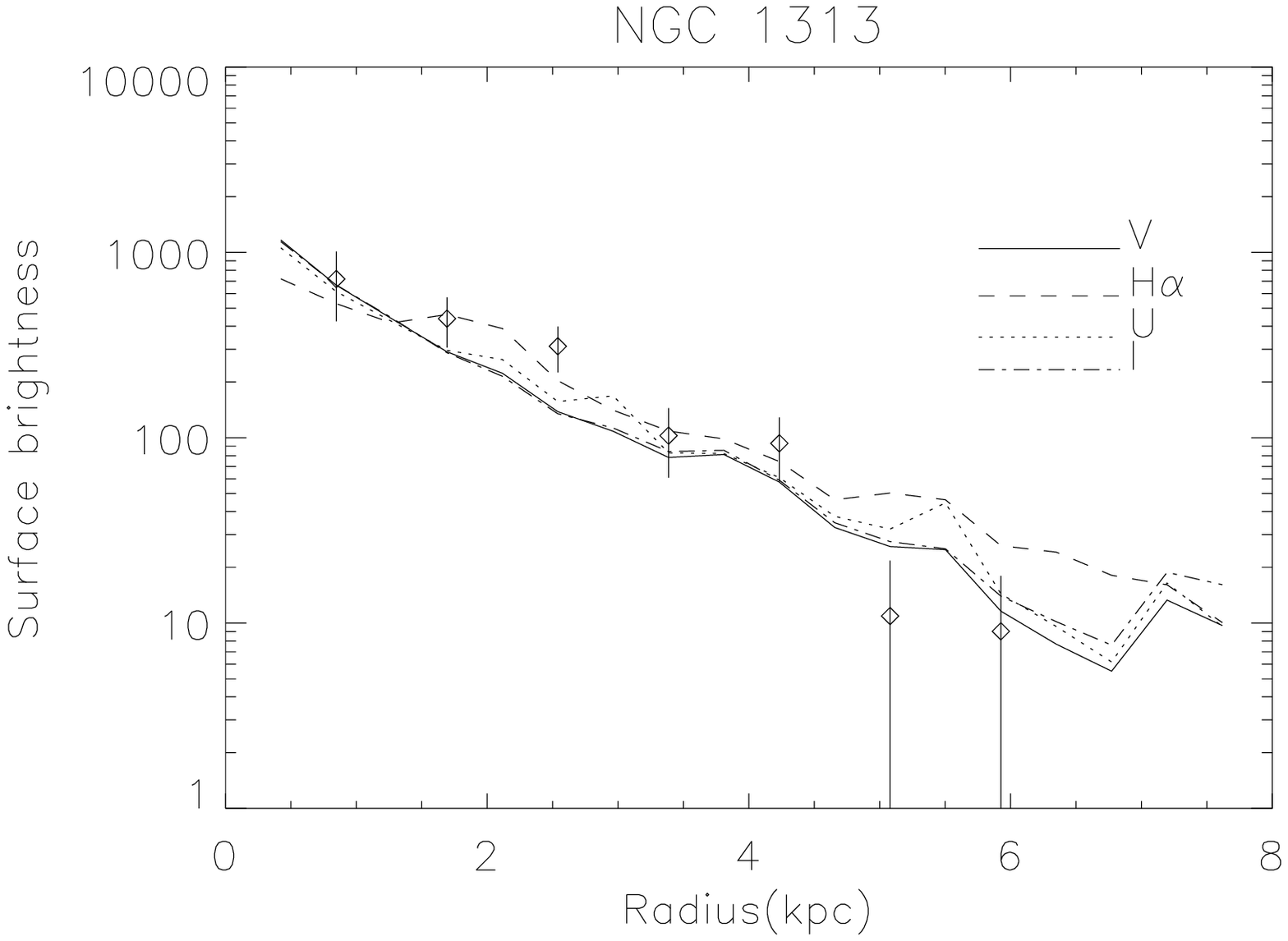}
\end{minipage}
\\
\begin{minipage}{8.5cm}
\epsfxsize=8.5cm
\epsfbox{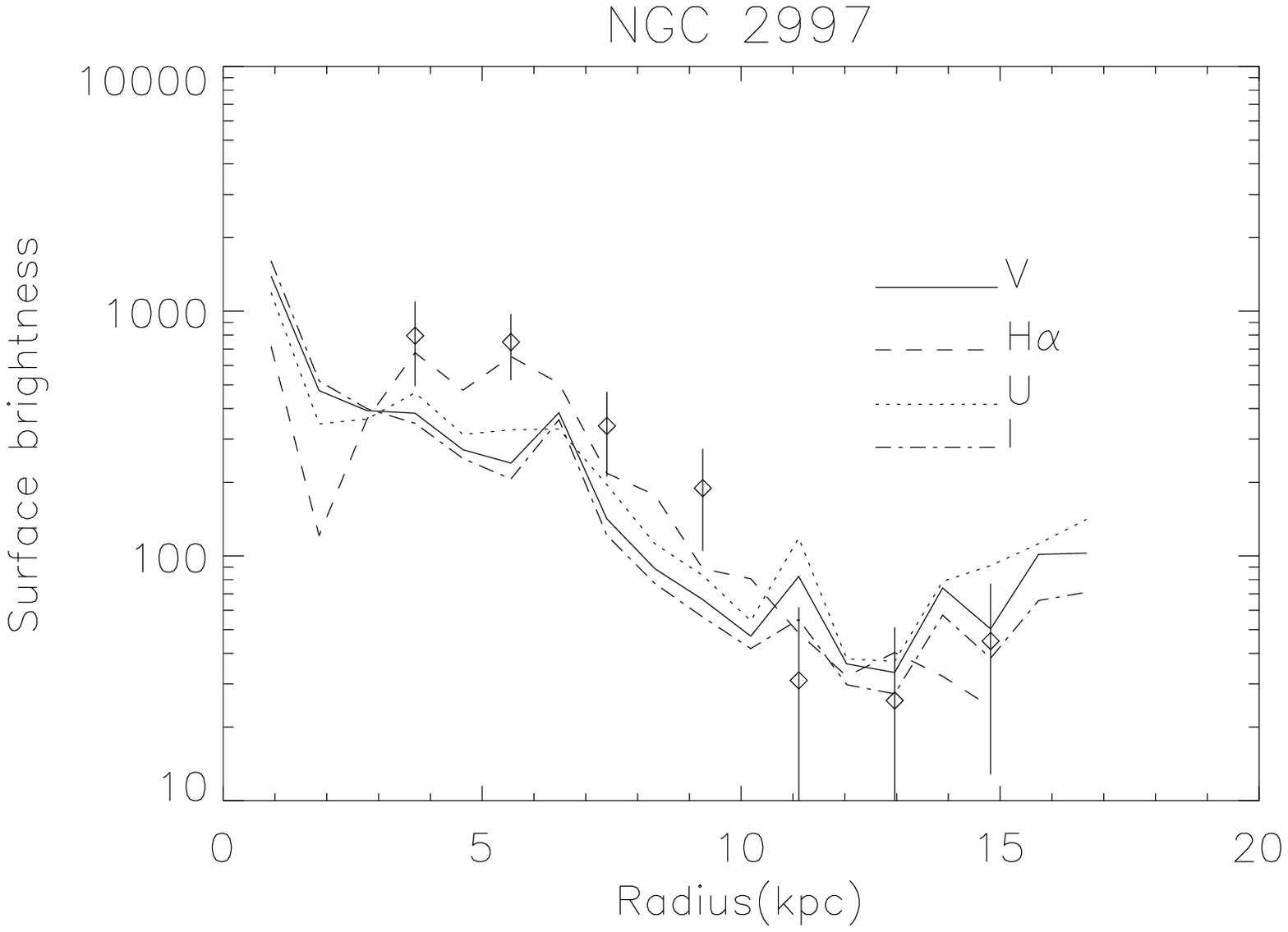}
\end{minipage}
\begin{minipage}{8.5cm}
\epsfxsize=8.5cm
\epsfbox{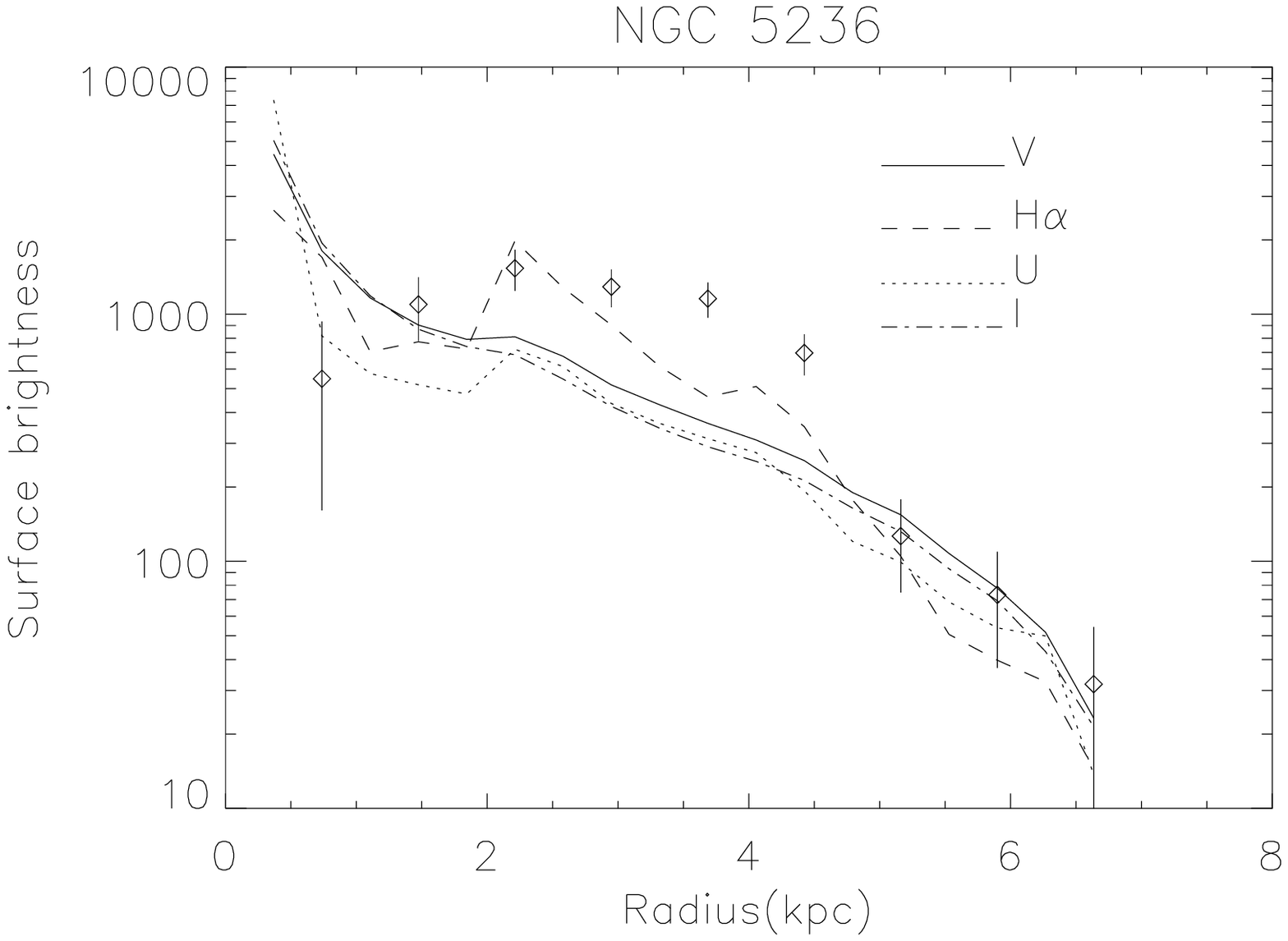}
\end{minipage}
\caption{Radial cluster distributions compared with surface brightness
profiles in $U$, $V$, $I$ and $H\alpha$. The dots with error bars show the
``surface density'' of clusters, and length of the error bars correspond to
poisson statistics.
\label{fig:rprof}
}
\end{figure*}

  As an attempt to investigate how cluster formation correlates with the
general characteristics of galaxies, we have compared the surface densities
of YMCs (number of clusters per unit area) as a function of galactocentric
radius with the surface brigthness in $U$, $V$, $I$ and $H\alpha$. Obviously,
such a comparison only makes sense for relatively rich cluster systems,
and is shown in Fig.~\ref{fig:rprof} for four of the most cluster-rich
galaxies in our sample. We did not include data for the apparently 
quite cluster-rich galaxy NGC~6946 in Fig.~\ref{fig:rprof} because of the
numerous Galactic foreground stars in the field of this galaxy which make
the cluster identifications less certain.

  The surface brightnesses were measured directly on our CCD images using 
the {\bf phot} task in DAOPHOT. In the case of $H\alpha$ we used 
continuum-subtracted images, obtained by scaling an R-band frame so that the 
flux for stellar sources was the same in the R-band and $H\alpha$ images, 
and subtracting the scaled R-band image from the $H\alpha$ image.
The flux was measured through a number of apertures with radii
of 50, 100, 150 $\ldots$ pixels, centered on the galaxies, and the background
was measured in an annulus with an inner radius of 850 pixels and a width
of 100 pixels. The flux through the i'th annular ring was then calculated
as the flux through the i'th aperture minus the flux through the (i-1)'th
aperture, and the surface brightness was finally derived by dividing with
the area of the i'th annular ring. No attempt was made to standard calibrate
the surface brightnesses, so the $y$-axis units in Fig.~\ref{fig:rprof}
are arbitrary.
  The cluster ``surface densities'' were obtained by normalising the
number of clusters within each annular ring to the area of the respective 
rings. Finally, all profiles were normalised to the V-band surface
brightness profile.

For all the galaxies in Fig.~\ref{fig:rprof} the similarity between the surface 
brightness profiles and the cluster surface densities is quite striking. In 
the cases of NGC~2997 and NGC~5236, where the $H\alpha$ profiles are markedly 
different from the broad-band profiles, the cluster surface densities seem 
to follow the $H\alpha$ profiles rather than the broad-band profiles.
Accordingly the presence of massive clusters must be closely linked with the
process of star formation in general in those galaxies where YMCs are present.
In order to get a complete picture one should include the clusters in the
central starbursts of NGC~2997 and NGC~5236, but this would, in any case,
affect the conclusions only for the innermost bin. 

\section{Discussion}

\begin{figure}
\epsfxsize=8.5cm
\epsfbox{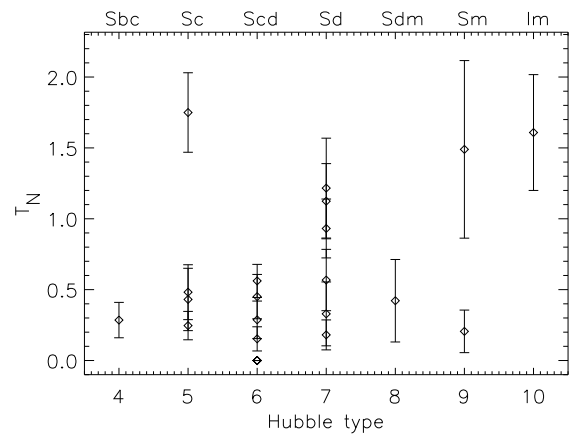}
\caption{$T_N$ values as a function of ``T''-type. 
\label{fig:t_tn}
}
\end{figure}

  Perhaps the most striking fact about the cluster-rich galaxies in our
sample is that they do not appear to have a lot of other properties in common. 
Fig.~\ref{fig:t_tn} shows the specific frequency $T_N$ as a function of
the ``T''-type (Table~\ref{tab:gal}), and does not support the
suggestion by Kennicutt \& Chu (1988) that the presence of YMCs in galaxies
increases along the Hubble sequence. Instead, at wide range of $T_N$
values is seen independently of Hubble type, so even if YMCs might be absent in
galaxies of even earlier types than we have studied here the phenomenon
cannot be entirely related to morphology.

  However, what characterises all these cluster systems is that 
they do not seem to have been formed during one intense burst of star 
formation. Instead, their age distributions as inferred from the ``S''
sequence are quite smooth (possibly with the exception of NGC~1313),
so in contrast to starburst galaxies like the Antennae or M82,
the rather ``normal'' galaxies in our sample have been able to maintain a
``production'' of clusters over a longer timescale, at least several
hundred Myr, in a more quiescent mode than that of the starburst galaxies.
The most luminous clusters we have found have absolute visual magnitudes
of about $M_V = -12$, about three magnitudes brighter than the brightest
open clusters in the Milky Way, but still somewhat fainter than
the $M_V = -13$ to $M_V = -15$ clusters in the Antennae and certain
starburst galaxies.

  One notable exception is NGC~6946 which is forming such a ``super star 
cluster'' just before our eyes. That cluster is located far away from 
the centre of the galaxy, something which is not unusual at all. Also in 
NGC~1313 the most massive cluster is located far from the centre of the host 
galaxy, at a projected galactocentric distance of about 3.7 kpc, and in the
Milky Way a number of high-mass (old) open clusters are found in the anticentre
direction, e.g. M67. It can of course not be excluded that a massive cluster 
like the one in NGC~6946 could be located in a region of the Galactic disk 
hidden from our view, but in any case the Milky Way does not seem to
contain any large number of young massive clusters as seen e.g. in NGC~5236
or NGC~1313.

  In general we find, however, that the distribution of YMCs follows the 
H$\alpha$ surface brightness profile, at least for those galaxies where the 
statistics allow such a comparison. Taking H$\alpha$ as an indicator of star 
formation, it then appears that in certain galaxies the formation of YMCs 
occurs whenever stars are formed. This raises the question whether the 
presence of massive cluster formation is correlated with global star formation 
indicators, such as $H\alpha$ luminosity or other parameters. These questions 
will be addressed in more detail in a subsequent paper (Larsen et. al. 1999).

  Two of the galaxies in our sample, NGC~5236 and NGC~2997, have a lot
of properties in common. Both galaxies are grand-design, high
surface-brightness spirals although 
NGC~2997 lacks the impressive bar of NGC~5236, and both were known to
contain massive star clusters near their centres also before this study.
We have identified rich cluster system throughout the disks of these two 
galaxies.

  In our opinion it is becoming clearer and clearer that a whole continuum 
of cluster properties (age, mass, size) must exist, one just has to look in 
the right places.   For some reason the Milky Way and many other
galaxies were only able to form very massive, compact star clusters during
the early phases of their evolution, these clusters are today seen as
globular clusters in the halos of these galaxies. Other galaxies
such as the Magellanic Clouds, M33 and NGC~2403 are able to form 
substantially larger number of massive clusters than the Milky Way even 
today, and in our sample of galaxies we have at least 5 galaxies that are
able to form clusters whose masses reach well into the interval defined by
the globular clusters of the Milky Way. Still more massive clusters are
being formed today in genuine starburst and merger galaxies such as
the Antennae, NGC~7252, M82 and others, and it seems that the masses
of these clusters can easily compete with those of ``high-end'' globular
clusters in the Milky Way. Whether YMCs will survive long enough to one 
day be regarded as ``true'' globular clusters is still a somewhat 
controversial question, whose definitive answer requires a detailed
knowledge of the internal structure of the individual clusters and
a better theoretical understanding of the dynamical evolution of star
clusters in general.

  One could also ask if the LF of star clusters really has an upper cut-off 
that varies from galaxy to galaxy, or if the presence of massive clusters is
merely a statistical effect that follows from a generally rich cluster
system. In order to investigate this question it is necessary to obtain
data with a sufficiently high resolution that the search for star clusters
can be extended to much fainter magnitudes than we have been able to do
in our study.

\section{Conclusions}

  The data presented in this paper demonstrate that massive star clusters
are formed not only in starburst galaxies, but also in rather normal 
galaxies. None of the galaxies
in our sample show obvious signs of having been involved in interaction
processes, yet we find that there is a large variation in the specific 
frequency $T_N$ of massive clusters from one galaxy to another. Some of 
the galaxies in our sample (notably NGC~1313 and NGC~5236) have considerably 
higher $T_N$ than the LMC, while other galaxies which at first glance could 
seem in many ways
morphologically similar to the LMC (e.g. NGC~300 and NGC~4395) turn out
to contain no rich cluster systems. In general there is no
correlation between the morphological type of the galaxies in our sample
and their $T_N$ values. Whether a galaxy contains massive star clusters or 
not is therefore not only a question of its morphology (as suggested by 
Kennicutt \& Chu 1988), so one has to search for correlations between
other parameters and the $T_N$ values. Within each of the galaxies that
contain populations of YMCs, the number of clusters as a function of
radius follows the H$\alpha$ surface brightness more closely than the
broad-band surface brightness, which implies that the formation of massive
clusters in a given galaxy is closely linked to star formation in general.

\begin{acknowledgements}
  This research was supported by the Danish Natural Science Research Council
through its Centre for Ground-Based Observational Astronomy.
This research has made use of the NASA/IPAC Extragalactic Database (NED)
which is operated by the Jet Propulsion Laboratory, California Institute
of Technology, under contract with the National Aeronautics and Space
Administration. We are grateful to J.V. Clausen for having read several 
versions of this manuscript, and the DFG Graduierten Kolleg 
``Das Magellansche System und andere Zwerggalaxien'' is thanked for
covering travel costs to S.S. Larsen.

\end{acknowledgements}

\end{document}